\documentclass[12pt,preprint]{aastex}
\usepackage{psfig,longtable}
\newcommand{\ASCA}{{\it ASCA}}
\newcommand{\Granat}{{\it Granat}}
\newcommand{\Einstein}{{\it Einstein}}

\newcommand{\XTE}{{\it XTE}}
\newcommand{\SAX}{{\it SAX}}
\newcommand{\Chandra}{{\it Chandra}}
\newcommand{\etal}{et al.}
\newcommand{\NH}{{$N_{\rm H}$}}
\newcommand{\NHUNIT}{${\rm H}\ {\rm cm}^{-2}$}

\newcommand{\FLUXUNIT}{${\rm erg}\ {\rm cm}^{-2}\ {\rm s}^{-1}$}

\newcommand{\LUMIUNIT}{${\rm erg}\ {\rm s}^{-1}$}
\newcommand{\GC}{Galactic Center}

\newcommand{\SgrAdiffuse}{the Sgr~A diffuse}
\newcommand{\SgrA}{Sgr~A}

\newcommand{\an}{Astron. Nach.}

\newcommand{\hms}[3]{$#1^{\rm h}#2^{\rm m}#3^{\rm s}$}

\lefthead{Sakano et al.}
\righthead{X-ray catalogue in the {\GC}}

\begin{document}

\title{{\ASCA} X-ray source catalogue in the {\GC} region}
\author{Masaaki Sakano\altaffilmark{1,2}}
\affil{Department of Physics and Astronomy,
        University of Leicester, University Road, Leicester  LE1 7RH, UK
 \email{sakano@cr.scphys.kyoto-u.ac.jp}}

\and

\author{Katsuji Koyama, 
 Hiroshi Murakami\altaffilmark{2} 
 }
\affil{Department of Physics, Kyoto University,
 Kyoto 606-8502 Japan,
 \email{koyama@cr.scphys.kyoto-u.ac.jp},
 \email{hiro@cr.scphys.kyoto-u.ac.jp}
 }

\and

\author{Yoshitomo Maeda\altaffilmark{2,3,4}}
\affil{Institute of Space and Astronautical Science (ISAS),
 Kanagawa 229-8510 Japan,
 \email{maeda@astro.psu.edu}}

\and

\author{Shigeo Yamauchi}
\affil{Faculty of Humanities and
 Social Sciences, Iwate University, 3-18-34 Ueda, Morioka, Iwate
 020-8550 Japan, 
 \email{yamauchi@iwate-u.ac.jp}}

\altaffiltext{1}{Space Utilization Research Program (SURP),
% Tsukuba Space Center (TKSC),
 National Space Develop Agency of Japan (NASDA),
 2-1-1 Sengen, Tsukuba, Ibaraki 305-8505 Japan}
\altaffiltext{2}{Research Fellow of the Japan Society for the Promotion of Science.}
\altaffiltext{3}{Subaru Telescope,National Astronomical Observatory of Japan, 650 North Aohoku Place, Hilo, HI 96720}
\altaffiltext{4}{Department of Astronomy and Astrophysics, The Pennsylvania State
University, 525 Davey Lab., University park, PA 16802-6305}

\begin{abstract}

The  {\ASCA} satellite made  107 pointing  observations on a  $5\times 5$ deg$^2$ region around the center of  our Milky Way Galaxy (the {\GC})  from 1993 to 1999.
In the X-ray images of the 0.7--3 keV or  3--10 keV bands, we found 52 point sources
and  a dozen diffuse sources. 
All the point sources are uniformly fitted with an absorbed power-law model.
For selected bright sources, Sgr~A$^*$, AX~J1745.6$-$2901, A 1742$-$294, SLX 1744$-$300, GRO~J1744$-$28, SLX 1737$-$282, GRS 1734$-$292, AX J1749.2$-$2725, KS 1741$-$293, GRS 1741.9$-$2853, and an unusual flare source XTE J1739$-$302, we present
further detailed spectral and timing analyses, and discuss their nature.
  The dozen extended X-ray sources comprise radio supernova remnants,
 giant molecular clouds, and some new discoveries. Most show emission lines from either highly ionized atoms or low-ionized irons.
 The X-ray spectra were fitted with either a thin thermal or power-law model.  
This paper summarizes the results and  provides the {\ASCA} X-ray source catalogue
in the {\GC} region.

\end{abstract}

\section{INTRODUCTION \label{sec:intro}}

   Stars, hot gas, and cold gas are densely distributed near the center of
 galaxies, which may lead to various activities, e.g.,  mass accretion on a massive black hole, starbursts, and supernova explosions.
  These phenomena are also suggested in the center of our Milky Way galaxy, the {\GC} region.
  Its proximity to us makes the {\GC} region an excellent  laboratory for  
 the most detailed study on such activity, hence it has been one of the major
 research objects (see reviews by Morris \& Serabyn 1996\markcite{Morris_S1996};
 Mezger, Duschl, \& Zylka 1996\markcite{Mezger1996}).

   Hard X-rays are key wavelengths  for the study of the {\GC} activity
because a putative massive black hole, starbursts, and their relics are all potential X-ray emitters,  and those X-rays are almost free from interstellar absorption.
Initially, hard X-ray studies have been made with non-imaging instruments (no mirror) (e.g., Kellogg {\etal} 1971\markcite{Kellogg1971};
Proctor {\etal} 1978\markcite{Proctor1978};
%Watson {\etal} 1981\markcite{Watson1981};
 Skinner {\etal} 1987\markcite{Skinner1987}; Kawai {\etal} 
1988\markcite{Kawai1988};
% Yamauchi {\etal} 1990b\markcite{Yamauchi1990b};
 Yamauchi {\etal} 1990\markcite{Yamauchi1990};
 Sunyaev {\etal} 1991b\markcite{Sunyaev1991asr};
 Sunyaev, Markevitch, \& Pavlinsky 1993\markcite{Sunyaev1993};
 Pavlinsky, Grebenev, \& {Sunyaev} 1994\markcite{Pavlinsky1994}; Churazov {\etal} 1994\markcite{Churazov1994}).

The most distinct components in the {\GC} region were 
variable point sources (e.g., Pavlinsky {\etal} 1994\markcite{Pavlinsky1994}),
 which are binaries with black holes or neutron stars, and the Galactic nucleus.
Other than the point sources,
 supernova remnants (SNRs) and largely extended high-temperature plasma were
 also found (e.g., Koyama {\etal} 1989\markcite{Koyama1989}).
  Most of the previous instruments, however, suffered from severe source-confusion problem
in the crowded {\GC} region.

{\ASCA} is the first satellite  equipped with
X-ray mirrors and detectors that have a high sensitivity and
reasonable  spectral and spatial resolutions in the wide energy band up to 10~keV, hence has been  used for  more detailed imaging  spectroscopy in the {\GC} region
 (e.g., Koyama {\etal} 1996\markcite{Koyama1996};
 Maeda \& Koyama 1996\markcite{Maeda_K1996};
 Maeda 1998\markcite{Maeda1998};
 Tanaka et al. 2000\markcite{Tanaka2000};
 Sakano 2000\markcite{Sakano2000}).
The next hard X-ray mirror satellite,  {\it Beppo-SAX}, also revealed
 new high-energy aspects near the {\GC} region (Sidoli {\etal} 1999\markcite{Sidoli1999}).

These works, however, have been limited to individual X-ray sources.
In order to obtain comprehensive and unbiased knowledge on 
the central 5$\times$5 degree$^2$ region, we have performed a complete  
survey since 1998 with {\ASCA}. 
This paper summarizes the results with the following order.
Details for the observations and the data screening are given in \S2.
In \S\ref{sec:image}, X-ray images of the {\GC} region in the
hard and soft energy bands are constructed.  Section~\ref{sec:res1:srcfind} provides 
the detected sources, their spectral parameters, and
identifications.  Further details for  selected
bright sources are presented in \S\ref{sec:selsrc}.
Throughout this paper, we assume the distance to the {\GC}
to be 8.5 kpc.

\section{OBSERVATIONS \label{sec:obs}}

   {\ASCA} has four X-ray telescopes (XRT) with focal plane detectors of
two Solid State Imaging Spectrometers (SIS0 and 1) and two Gas Imaging
Spectrometers (GIS2 and 3).  Instrument details are
found in Serlemitsos {\etal} (1995)\markcite{Serlemitsos1995}, Burke {\etal} (1991\markcite{Burke1991}, 1994\markcite{Burke1994}),
Yamashita {\etal} (1997)\markcite{Yamashita1997}, Ohashi {\etal} 
(1996)\markcite{Ohashi1996}, Makishima {\etal}
(1996)\markcite{Makishima1996}, while a general description of {\ASCA} can be found in Tanaka, Inoue, \& Holt (1994)\markcite{Tanaka1994}.

   Since the field of view of SIS is smaller than that of GIS, and
 the performance of the SISs has been significantly degraded since 1995
 due to the radiation of charged
 particles \citep{Dotani1997}, this paper refers to the GIS data unless otherwise stated.
  The GISs were generally operated in the PH mode with
the standard bit-assignment (10-8-8-5-0-0; see Ohashi {\etal} 
1996\markcite{Ohashi1996}), unless otherwise noted (Table~\ref{tbl:obs:log}).  

\begin{figure}[htbp]
 \centering
 \mbox{\plotone{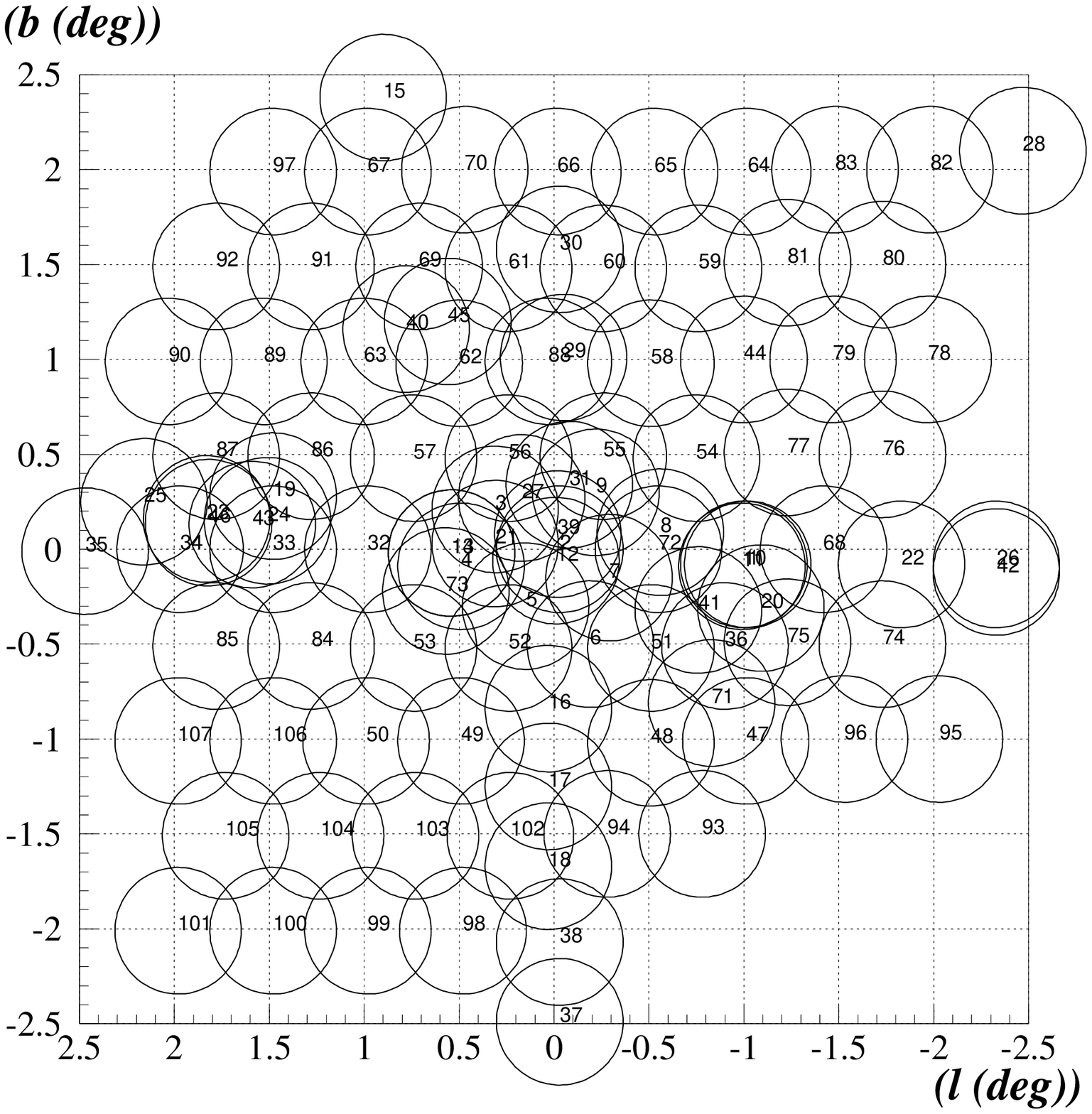}}
 \caption[f1.eps]{
   The GIS fields of view of all the {\ASCA} pointings in the {\GC}
 $5^{\circ}\times 5^{\circ}$ region with the galactic
 coordinates.  The radius of each circle is 20$'$.
  See Table~\ref{tbl:obs:log} for the field IDs (Obs-IDs).
 \label{fig:obs:fov}}
\end{figure}

The GIS fields at selected targets including 4 TOO (Target Of Opportunity) transient sources
 were observed with the pointing mode.  In order to fill blank fields, 
we performed follow-up survey observations from 1998 to 1999.
In total, the {\GC} $5^{\circ}\times 5^{\circ}$ region was
observed 107 times from 1993 to 1999.  Table~\ref{tbl:obs:log}
and Figure~\ref{fig:obs:fov} summarize the observation log with the
sequence numbers (Obs-ID, here and after).
The total exposure time in the {\GC} $5^{\circ}\times
 5^{\circ}$ area is about 1600 ksec, however, the exposure for each GIS field
scatters largely from 5 to 200 ksec, depending on the observation 
modes or  objectives (targets). 
For the follow-up survey fields, shorter exposures (e.g., 10 ksec) were generally allocated,
 while longer exposures (e.g., 200 ksec) were achieved for multiple-pointings
on selected target fields.

%\subsection{Data screening}

Data screening was performed, using the standard method  described in 
Day {\etal} (1995)\markcite{Day1995}.   
The data with  telemetry saturation of larger than 2$\sigma$ level
 are discarded, except for Obs-ID  27 field (see 
Table~\ref{tbl:obs:log}),  where the telemetry  saturation were corrected 
according to Nishiuchi {\etal} (1999)\markcite{Nishiuchi1999}.
The rise-time discrimination was  applied to reject non-X-ray 
background (NXB), except for Obs-ID$=$27 (see Table~\ref{tbl:obs:log}).
For timing analyses,  barycentric arrival time correction was  made for all the X-ray 
photons.

\placetable{tbl:obs:log}
\placefigure{fig:obs:fov}

\section{X-RAY IMAGES \label{sec:image}}

\begin{figure}[htbp]
 \centering
% \psfig{file=imghard.eps,width=\textwidth,clip=}
%%(a) 3.0--10.0 keV band.
%\vspace*{0.5cm}
 \mbox{\plottwo{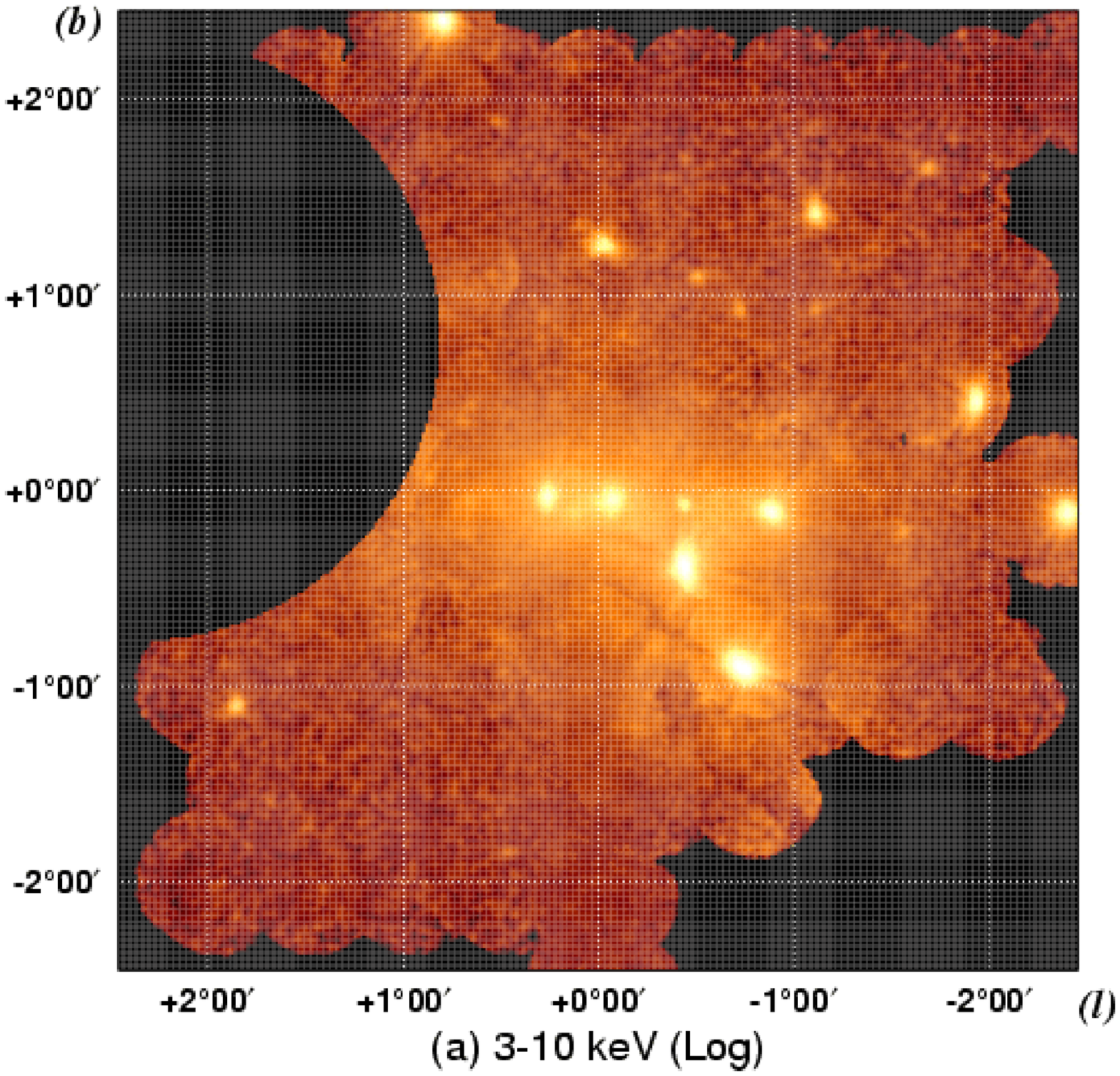}{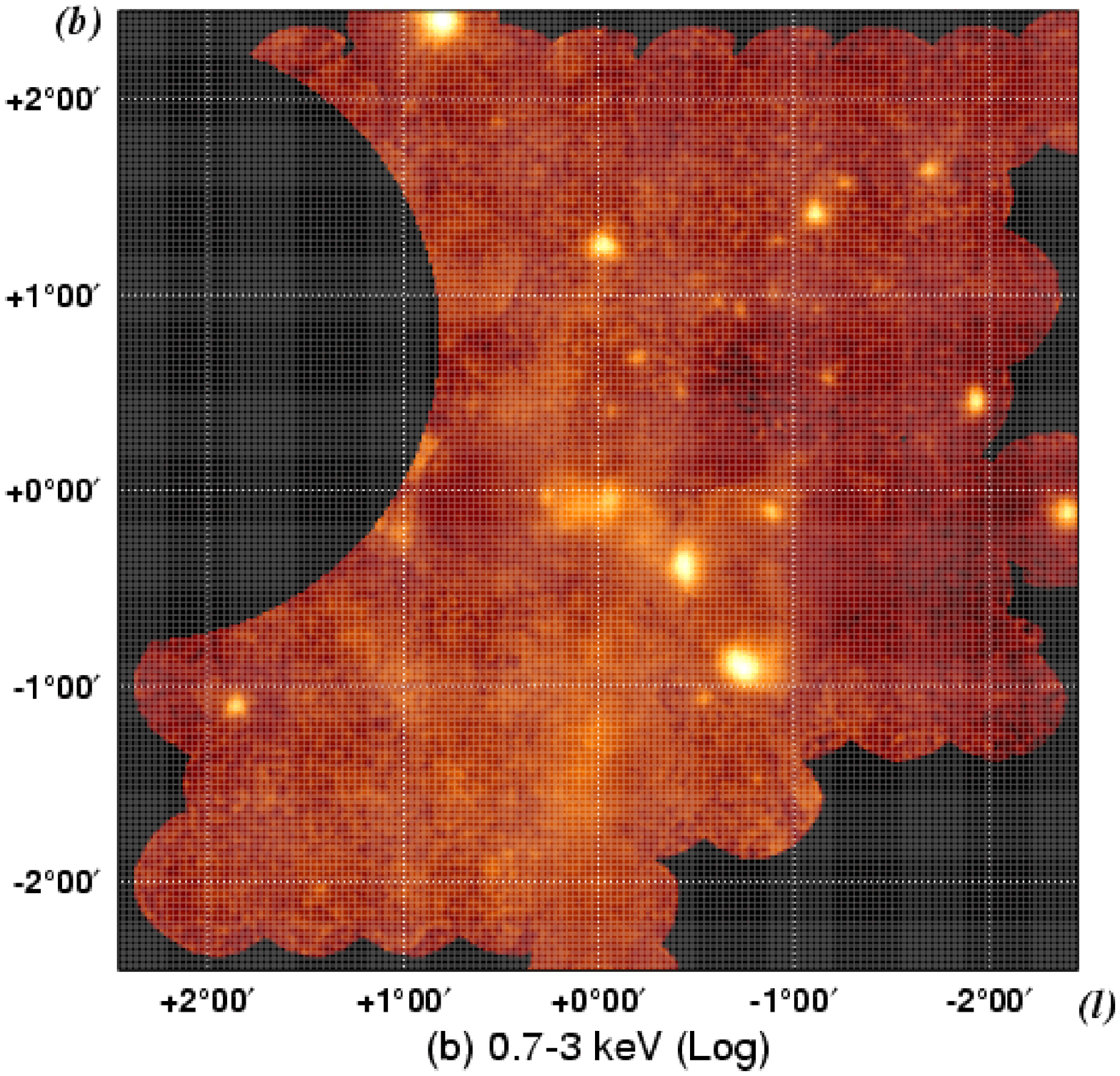}}
 \caption[f2a_col.eps,f2b_col.eps]
{{\ASCA} X-ray images in the {\GC} $5^{\circ}\times 5^{\circ}$ region
 for the energy bands of (a) 3.0--10.0 and (b) 0.7--3.0 keV
 with the galactic coordinates.
 The color levels are logarithmically spaced.
  The data of GIS2 and GIS3 are summed,
 smoothed with a Gaussian filter of $\sigma =$ 3 pixels
 ($\sim$ 0.75 arcmin),
 and corrected for exposure, vignetting, and the detection efficiency
 with GIS grid,
 after non-X-ray background is subtracted, according to the method
 described in Sakano (2000)\markcite{Sakano2000}.
   A region around $(l_{\rm II}, b_{\rm II}) = (2^{\circ}\!.3, 0^{\circ}\!.8)$
 is heavily contaminated by the stray light from GX~3+1, hence is deleted
 as is seen by a blank semi-circle.
 \label{fig:res1:mosaic-img}}
\end{figure}

Figures~\ref{fig:res1:mosaic-img}a and \ref{fig:res1:mosaic-img}b show the X-ray images in the {\GC}
$5^{\circ}\times 5^{\circ}$ region with the energy bands of 3.0--10.0 and
0.7--3.0 keV, respectively.  A blank region around $(l_{\rm II}, b_{\rm II}) = (-1^{\circ}\!.5, -1^{\circ}\!.5)$ is
 due to  unexpected accidents in the {\ASCA} operation.
   A region around the brightest source GX3+1 is heavily contaminated by the stray light (see next paragraph), hence is deleted as is seen by a blank  semi-circle around $(l_{\rm II}, b_{\rm II}) = (2^{\circ}\!.3, 0^{\circ}\!.8)$.
   For the same reason, some pointing data for extremely bright transient
 sources, such as the flare data of GRO~J1744$-$28, are discarded.

In the hard energy band of 3--10 keV (Fig.~\ref{fig:res1:mosaic-img}a),
the brightest are X-ray point sources.  
Radial structures seen around  ($l_{\rm II}$, $b_{\rm II}$) = 
 ($-0^{\circ}\!.2$,
$-0^{\circ}\!.8$) and ($0^{\circ}\!.6$,
$1^{\circ}\!.0$) are artifacts made by  the stray lights from the bright sources.  
Since the stray lights are time variable, some artificial 
discontinuity cannot be removed in the re-constructed image, or mosaic-image (e.g., 
Serlemitsos {\etal} 1995\markcite{Serlemitsos1995}).  

Ignoring  the artificial structures around bright point sources, we still see
further diffuse structures, in particular near the Galactic plane.  
These are local enhancements of the hot plasma and X-ray reflection nebulae 
(Koyama {\etal}
1996\markcite{Koyama1996}; Maeda 1998\markcite{Maeda1998}; Maeda {\etal} 
1999\markcite{Maeda1999}; Murakami {\etal} 1999\markcite{Murakami1999}, 
2000a\markcite{Murakami2000a}, 2000b\markcite{Murakami2000b}, 
2001\markcite{Murakami2001}; Sakano {\etal}
1999c\markcite{Sakano1999an}, 2000a\markcite{Sakano2000diffuse}; Tanaka {\etal} 2000\markcite{Tanaka2000}).

   The X-ray image in the soft energy band (0.7--3 keV;
 Fig.~\ref{fig:res1:mosaic-img}b) is largely 
different from  that of the hard energy band.  
The bright hard-band point sources, located  near the Galactic plane,
become  fainter due to the  heavy absorption (e.g., Sakano {\etal}
1999b\markcite{Sakano1999asr}; Sakano 2000\markcite{Sakano2000}).
Diffuse clumpy  structures are clearer than that in the hard band.
   The stray light structures
 become dim, due simply to the reduced flux of the  original
point sources.

\placefigure{fig:res1:mosaic-img}

\section{X-RAY SOURCE LIST IN THE GALACTIC CENTER REGION \label{sec:res1:srcfind}}

Since the diffuse  background  and stray light  structures are complex,
source detection cannot be a simple  job.   We therefore developed the source detection procedures following the method by Ueda {\etal} (1998\markcite{Ueda1998}, 1999\markcite{Ueda1999}).  

We first pick up source candidates from each pointing image.  
This procedure can resolve neighboring sources separated  as small as 1 arcmin.
We then determine the accurate source positions by a 2-dimensional
image-fitting of the raw GIS2$+$3 image, using  the relevant point
spread function (PSF) and the background taken from the off-plane
 blank sky.   The normalization of the background is allowed to be free.
We apply the fitting to narrow regions with the size of
$10'\times 10'$ centered at the source to minimize the spatial fluctuation
of the background in the fitting region.
These source finding procedures are applied separately to the soft-band
(0.7--3.0 keV) and the hard-band (3.0--10.0 keV) images of each observation.

Although the statistical errors of the positions
are  less than 10$''$,  the systematic errors due to the background 
fluctuation and the calibration uncertainty is far  larger,  particularly near the edge of the GIS field.
 Thus, the nominal positional error radius is 50$''$
 in 90\% confidence.

For each of the candidate sources, we accumulate the spectrum from
a circle with a radius of 3$'$.  The background spectrum is made from the annular region with radii of 3$'$--5$'$ centered at the source position.  In these accumulations, we reject the pixels located within 2$'$ from nearby source candidates or diffuse enhancements. 
We define the significance to be the ratio of the source counts 
 to the 1$\sigma$ fluctuation of the relevant background,
and  set the  source detection criteria to be at the 5$\sigma$ level.

Above procedures occasionally picked up ``false'' sources from  bright diffuse structures, either real diffuse  or  stray light,  as  candidates of a point source  or superposition of point sources.  We therefore inspected  all the source images and removed  the ``false'' candidates from the source catalogue.

  Each source spectrum was fitted with an absorbed power-law function.  When the 
spectral parameters were not constrained,  we used the best-fit values obtained with 
the best-quality spectrum (for multiple pointing) or fixed them to the lower-limit or 
upper-limit values, and estimated the 0.7--10 keV flux. 
All the results are  summarized  in Table~\ref{tbl:res1:src-list}.

%%\section{Identification of bright sources}

  The source identification and the counterpart search were made, using the
past references and the SIMBAD database. 
 Thirty-one sources are found to have a counterpart. The results are  listed in
Table~\ref{tbl:res1:src-list}.

By the inspection of the X-ray image and discarding the stray light structures,
 we found  a dozen diffuse X-ray sources.  Most of them are X-rays from radio SNRs, but some are new discoveries.  The spectra are fitted with an absorbed  thin thermal plasma model
 or a power-law function (and emission lines).
  The best-fit parameters and other relevant information are summarized in Table~\ref{tbl:res1:extended}.

\placetable{tbl:res1:src-list}
\placetable{tbl:res1:extended}

\section{NATURE OF SELECTED BRIGHT SOURCES \label{sec:selsrc}}

\subsection{Sgr~A$^*$ \& AX~J1745.6$-$2901 \label{sec:sgra}}

 {\ASCA} has observed the {\SgrA} region four times (see Table~\ref{tbl:res1:src-list}):
 twice in 1993 (Obs-ID=2, 7), once in 1994 (Obs-ID=12) and in 1997
(Obs-ID=39). The observation span is thus 3.5~yrs, which is the longest
monitoring of {\SgrA} ever achieved by hard X-ray imaging instruments.
We thus study long-term behaviors of X-ray sources in the {\SgrA}
region.  Since
\citet{Koyama1996} reported that the spatial structure near to {\SgrA} 
is highly complicated, we analyze the data taken with the SIS
detector, of which the spatial resolution is better than that of
GIS. The SIS data are available in three of the four observations
 (Obs-IDs of 2, 12, and 39).  Since the detailed results about the 1993 and
 1994 data were already reported by \citet{Maeda1996} and \citet{Koyama1996},
 we mainly concentrate on the 1997 data unless otherwise stated.

\begin{figure}[htbp]
 \centering
 \mbox{\plotone{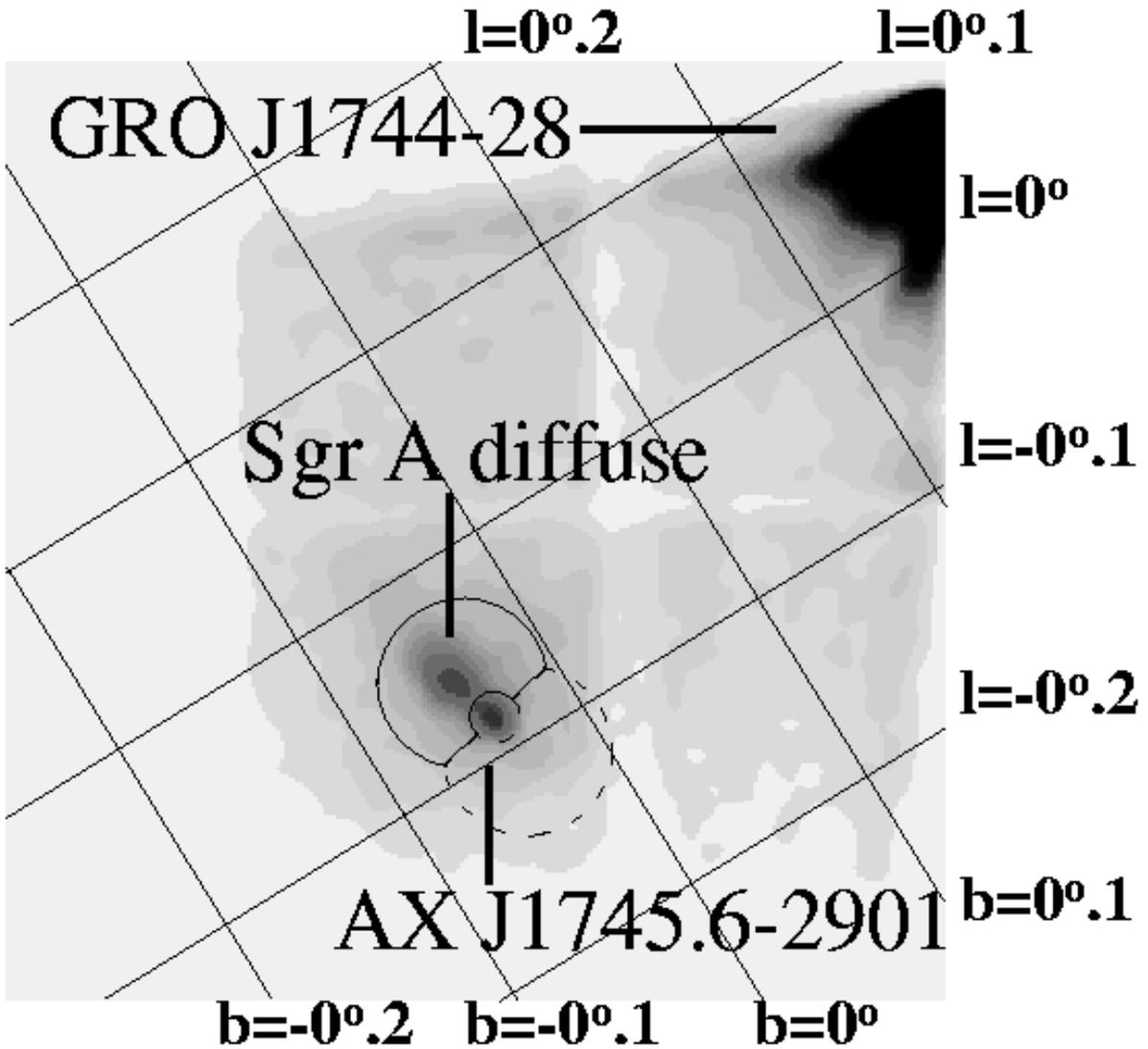}}
 \caption[f3.eps]{
   X-ray image of the Sgr~A region for the 3--10 keV band in 1997 (Obs-ID=39)
 taken with the SIS 4-CCD mode.
  X-ray photons for the spectra of the {\SgrA} diffuse and the background are
 accumulated from the region encircled by solid and dashed lines, respectively.
   The central small circle
 with a radius of 0$'\!$.8 is not used for the source- or the background-spectrum
 for the {\SgrA} diffuse emission, but for the light-curve
 of AX J1745.6$-$2901.
 \label{fig:sgra-img}}
\end{figure}

\placefigure{fig:sgra-img}

\begin{figure}[htbp]
 \centering
 \epsscale{0.9}
 \mbox{\plotone{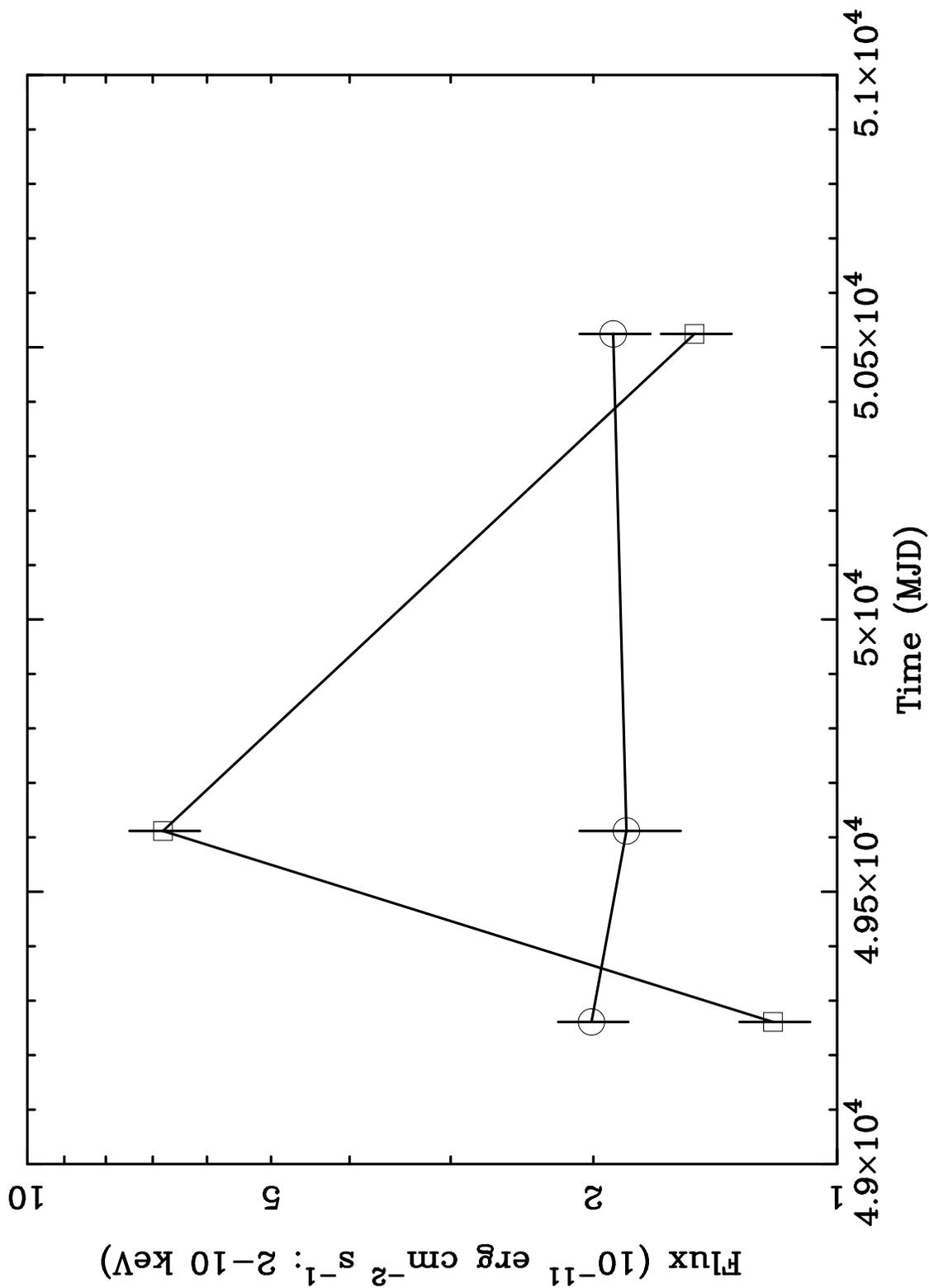}}
 \caption[f4.eps]{
   Flux histories of the Sgr~A region (open circle) and AX~J1745.6$-$2901
 (open rectangle), observed with {\ASCA}.  The flux of the Sgr~A region
 includes that of the Sgr~A diffuse plasma and possible contribution
 from Sgr~A$^*$.
 \label{fig:sgra-flux}}
\end{figure}

   Figure~\ref{fig:sgra-img} shows an SIS image in 1997 for the 3--10 keV band.
 Two X-ray sources seen in the 1993 and 1994 data
 \citep{Koyama1996,Maeda1996} also appeared in 1997; the north-east source
 (here and after, the {\SgrA} diffuse) is slightly extended ($\sim$ a few
arcmin), of which the spatial peak is consistent with {\SgrA}*, while
 the south-west source is a low mass X-ray binary AX~J1745.6$-$2901, named by
\citet{Kennea_S1996}. The {\SgrA} diffuse showed no flux variation (see \S\ref{sec:sgra-diffuse})
 while AX J1745.6$-$2901 was variable; the persistent fluxes of AX J1745.6$-$2901
 in the 2--10 keV band were 1.2, 6.8, and
$1.5\times10^{-11}$ {\FLUXUNIT} in 1993, 1994 and 1997,
 respectively.  We thus found that AX J1745.6$-$2901 was
 in the low-flux state in 1993 and 1997, and in the high
 state in 1994.
   Figure~\ref{fig:sgra-flux} summarizes the variabilities of both the sources.

\placefigure{fig:sgra-flux}

\subsubsection{AX J1745.6$-$2901 \label{sec:axj1745}}

\begin{figure}[htbp]
 \centering
 \epsscale{0.8}
 \mbox{\plotone{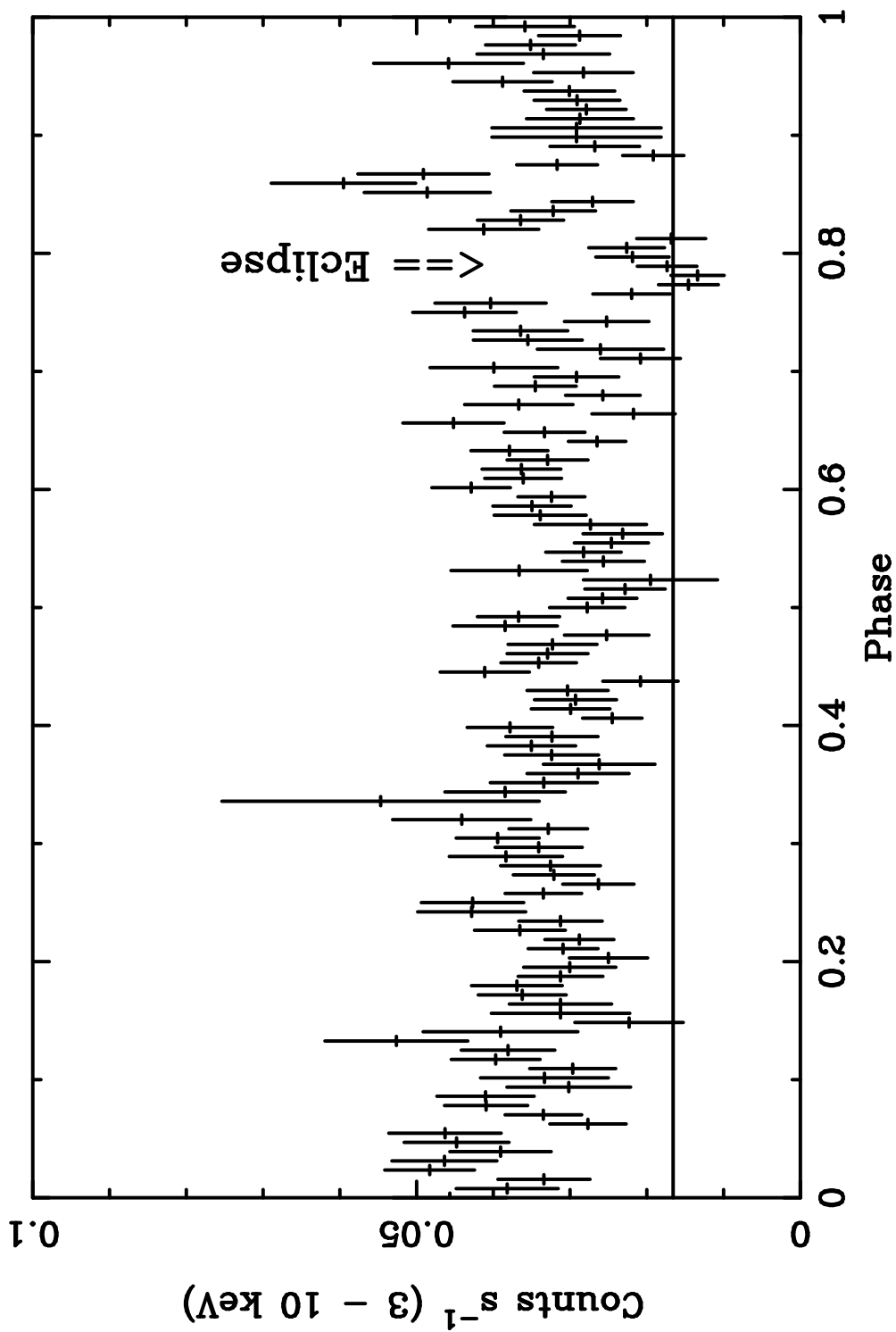}}
 \caption[f5.eps]{
   Folded light curve (3--10 keV) of AX J1745.6$-$2901 with the period
 of 8.356~hr in 1997 (Obs-ID=39) taken with SIS.  The phase 0 epoch
 is set to MJD 50523.0.
 The solid line gives the background level.
 A possible dip corresponding to the eclipse can be seen.
 \label{fig:axj-lcur-1997}}
\end{figure}

   \citet{Maeda1996} discovered intensity dips with the interval of
 $8.356\pm0.008$~hr due to the eclipse from this source in the 1994 data.
  We newly analyze the 1997 data.  We extracted X-ray photons from a circular
 region with a radius of 0$'$.8, and made a light curve, just the same as
 \citet{Maeda1996}.   Then, we folded the light curves with 8.356~hr,
 and found a possible dip also in 1997 (Fig.~\ref{fig:axj-lcur-1997}).
  Note that non-detection of the dip in 1993 is probably due to its short
 exposure of 17~ksec, which is about one-quarter of that in 1997 (70~ksec).

To constrain the orbital period, we determined the center of the
 eclipsing phase to be MJD 49610.2326(2) and MJD 50523.275(1) in the
 1994 and 1997 observations, respectively, by assuming 
 a well-type light curve for the eclipse.
  The time interval between
these two eclipse-centers is then 913.043(1) days.  When no change
 of the orbital period is assumed, this time interval should be $P \times N$,
where $P$ is the orbital period and $N$ is an integer. From the period
of $8.356\pm0.008$ hr constrained by \citet{Maeda1996},
 we derived more accurate orbital
period to be either 8.34782, 8.35100, 8.35419, 8.35737, 8.36056 or
8.36375 hrs with each error of 0.00001 hr.

% Note that no burst was detected in the low state in 1997, the same as in 1993,
%  while an X-ray burst event was observed in 1994 \citep{Maeda1996}. 

\placefigure{fig:axj-lcur-1997}

\subsubsection{The {\SgrA} Diffuse \label{sec:sgra-diffuse}}

   We made the SIS spectra of {\SgrAdiffuse} for the three observations.
 All of the data in the 1993 observation (Obs-ID=2) were available, whereas those
for the 1994 and 1997 observations were not.  For the 1994
observation, only the data during the eclipse phase of AX J1745.6$-$2901
 were used, because X-rays from AX J1745.6$-$2901 in non-eclipse phase,
 of which the flux was an order of magnitude larger than that of {\SgrAdiffuse},
 highly contaminated the {\SgrA} diffuse spectrum.
  For
 the 1997 observation, we used only the 1-CCD-mode data, in which the energy
 resolution was much better than that in the 4-CCD-mode
\citep{Yamashita1997}. The effective exposure time is then reduced to
be 18, 2.5, 37 ksec for the 1993, 1994, and 1997 data, respectively.

For all the three observations, the spectra of the {\SgrA} diffuse and
the background were taken from the paired regions given
 in Figure~\ref{fig:sgra-img}.  By subtracting the background data, we
 removed spill-over X-rays from {AX J1745.6$-$2901} and another transient 
source
GRO J1744$-$28 only seen in 1997.

\begin{figure}[htbp]
 \centering
 \epsscale{0.8}
 \mbox{\plotone{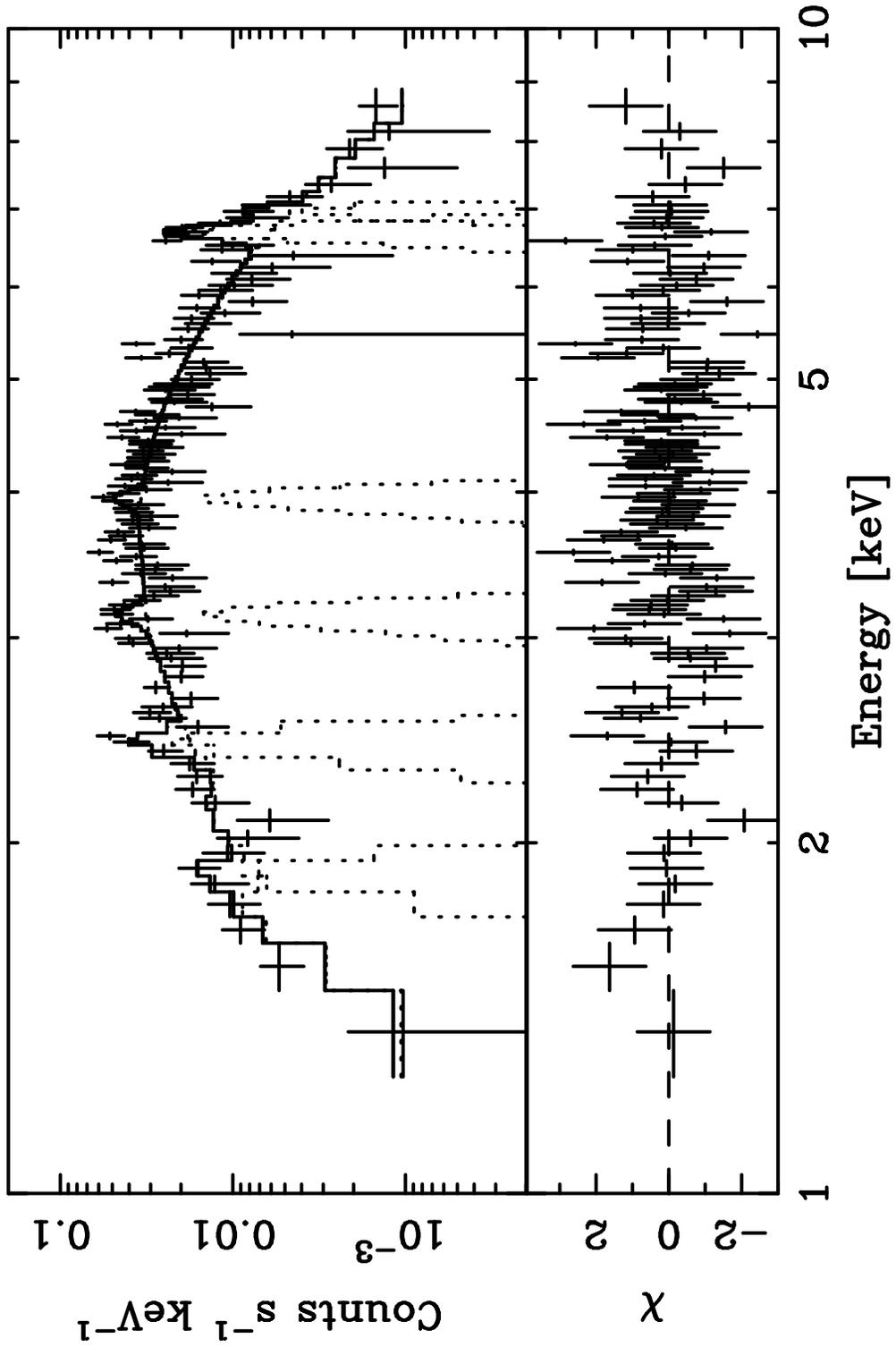}}
 \caption[f6.eps]{
   The SIS spectrum of the {\SgrA} diffuse in 1993 (Obs-ID=2).
 The spectrum is fitted with a model of bremsstrahlung and Gaussian lines.
 \label{fig:sgra_spec}}
\end{figure}

   Figure~\ref{fig:sgra_spec} shows the background-subtracted spectrum
 in 1993, which has by far the best quality in these observations.
  We fitted the spectrum in 1993 with a model of
 bremsstrahlung and Gaussian lines. From the line energies, we can identify
 these lines as K$\alpha$ transitions of highly ionized ions of silicon,
 sulfur, argon, calcium, and iron atoms.
  We then fitted the spectrum again, fixing the center energies of
 each line to the theoretical values, and determined the line fluxes.
   The best-fit parameters are listed in Table~\ref{tbl:sgra_spec}.

  Thus, we quantitatively confirm the existence of optically thin thermal
 plasma just surrounding {\SgrA}*, which was initially reported
 by \citet{Koyama1996}.
  The best-fit electron temperature is as
high as $\sim$8~keV, which is consistent with those reported by
\citet{Koyama1996} and \citet{Sidoli_M1999}. 
  For the {\GC} plasma extending over the 1-degree square region,
  \citet{Koyama1996} reported the prominent 6.4- and 6.97-keV lines,
 which are K$\alpha$ emission of cold and hydrogen-like iron, respectively.
 Therefore, we included these two lines in the fitting,
 but obtained no significant line fluxes (Table~\ref{tbl:sgra_spec}).

In order to study flux variations of the {\SgrA} diffuse, we fitted the
spectra taken in 1994 and 1997 with the same model given in
 Table~\ref{tbl:sgra_spec}, allowing only the global normalization factor
to be adjusted.  The model well reproduced the spectral shapes of the
two spectra.
The best-fit fluxes of the three observations were constant
within the statistical error, $\sim$2$\times$10$^{-11}$ erg~s$^{-1}$ cm$^{-2}$ (2--10 keV). 
 Hence, the flux and the spectral
shape of {\SgrAdiffuse} should be steady in the long time-span of the
three observations (Fig.~\ref{fig:sgra-flux}).

\placefigure{fig:sgra_spec}
\placetable{tbl:sgra_spec}

\subsubsection{Sgr A* \label{sec:sgra-star}}

The {\SgrA} diffuse, peaked at {\SgrA}*, is dominated by diffuse
emission. With the three observations, {\SgrAdiffuse} showed no
 time-variability, which suggests that {\SgrA}* emits no significant point-like
 X-ray; the conservative upper-limit of the X-ray flux of {\SgrA}*
  is
 $F_{\rm X, {\SgrA}^*} < 2 \times 10^{-11}$ {\FLUXUNIT} (2--10~keV).
  The X-ray luminosity of {\SgrA}* corrected for the
 interstellar absorption of 7$\times$10$^{22}$ H cm$^{-2}$ is then calculated
 to be
 $L_{\rm X, {\SgrA}^*} < 3\times10^{35}$ {\LUMIUNIT} (2--10~keV).
This upper limit is consistent with that measured with {\SAX}
\citep[$3\times 10^{35}$ erg s$^{-1}$;][]{Sidoli_M1999}
 and the positive flux recently reported 
with {\Chandra} \citep[$2\times 10^{33}$ erg s$^{-1}$;][]{Baganoff2001}.

Many hard X-ray observations near to {\SgrA}* with the non-imaging
instruments had been performed before {\ASCA} observations \citep[for
a summary of these observations, see][]{Maeda1996}.  These
observations suggested a long-term time-variability of {\SgrA}* with
the 3--10 keV flux ranging (1--16) $\times$10$^{-11}$ erg s$^{-1}$
cm$^{-2}$, which corresponds to the absorption-corrected luminosity of
about (1--20) $\times$10$^{35}$ erg s$^{-1}$ \citep{Skinner1987,Kawai1988,Sunyaev1991sov}.
 Our {\ASCA} monitoring over four years found that the low 
mass X-ray binary AX J1745.6$-$2901 is always present, which supports the 
argument by \citet{Maeda1996} that the X-ray fluxes previously reported with
 non-imaging instruments might have suffered from possible
 contamination of this source.  If the flux-change
 were due to AX J1745.6$-$2901, the periodic eclipse should be detected
 when the {\SgrA} region is in high state. 
 However, ART-P reported no eclipse in the high state \citep{Sunyaev1991sov};
and accordingly
the variability would not be likely due to AX J1745.6$-$2901, but possibly
due to {\SgrA}* (R. Sunyaev, private communication). In fact, the first
{\Chandra} observation reports a hint of variability for a small flare
 from {\SgrA}* \citep{Baganoff2001} although the flux level is an order of $10^{33}$ 
erg~s$^{-1}$. Presence of the moderately large variability should be
tested by future {\Chandra} and {\it XMM-Newton} observations.

\subsection{A 1742$-$294, SLX 1744$-$300 \& GRO~J1744$-$28 \label{sec:bursters}}

We found 11 X-ray bursts from A~1742$-$294. The X-ray features are characteristic to the type-I bursts, hence we confirmed the previous results (e.g., Lewin {\etal} 1976\markcite{Lewin1976})

We clearly resolved SLX 1744$-$300 from a nearby burster, SLX 1744$-$299 \citep{Pavlinsky1994} separated by 3$'$. Although we found 11 bursts from SLX 1744$-$300, no burst was detected from SLX 1744$-$299.
SLX 1744$-$300 is known to  exhibit X-ray bursts with complicated
profiles (Skinner {\etal} 1990\markcite{Skinner1990_300}, 
Pavlinsky {\etal} 1994\markcite{Pavlinsky1994}). The detailed natures of the bursts, however, were  not studied with {\ASCA} due to the limited statistics.  

GRO J1744$-$28  is a  type-II burster and a 0.47~s pulsar  
with an orbital period of 11.83 days
(e.g., Finger {\etal} 1996\markcite{Finger1996};
Kouveliotou {\etal} 1996\markcite{Kouveliotou1996}).
Detailed natures with the {\ASCA} observations on and before 1997 March
are found in Nishiuchi {\etal} (1999)\markcite{Nishiuchi1999}.
This source was observed again on 1998 September 7 (Obs-ID=55 \& 56) with {\ASCA}, but
 no significant flux was detected
 with the 3$\sigma$ upper limit of 2$\times 10^{-12}$ {\FLUXUNIT}.

\subsection{XTE J1739$-$302 \label{sec:xtej1739}}

A transient source XTE J1739$-$302 was discovered with {\XTE} at (\hms{17}{39}{00}, $-$30$^{\circ}$16$'\!$.2 (J2000)) on 1997 August 12
(Smith {\etal} 1997\markcite{Smith1997}, 1998b\markcite{Smith1998_1739}).  Although it was the brightest source in
the {\GC} region on August 12 ($\sim 3.0\times 10^{-9}$ {\FLUXUNIT}
from 2 to 25 keV),  it was below the detection limit on  9 days before and 2 days after the flare. The {\SAX}/WFC observation made on 1997 September 6 \citep{Smith1998_1739} found no flux from this source.

\begin{figure}[htbp]
 \centering
 \mbox{\plotone{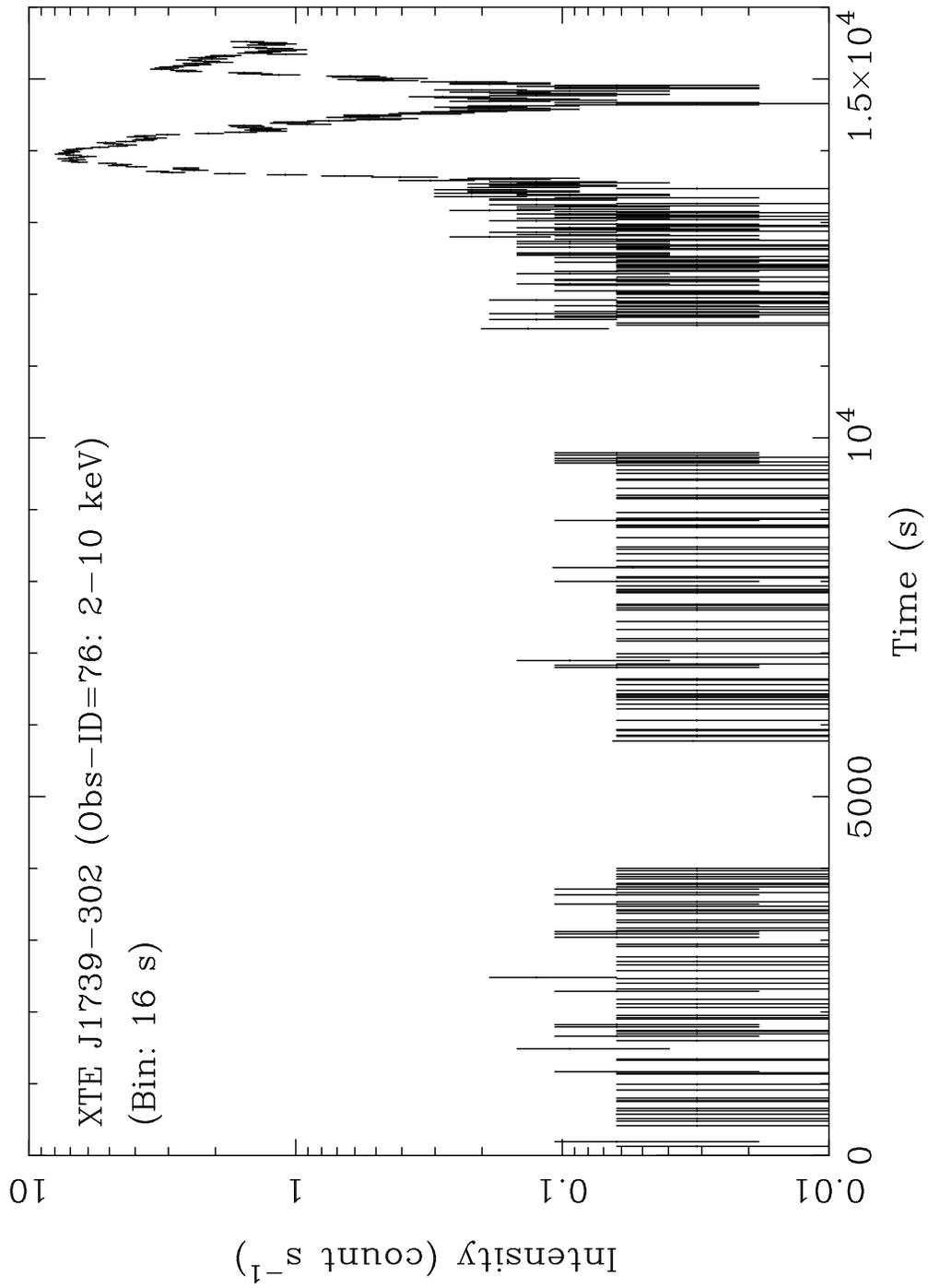}}
 \caption[f7.eps]{
   Light curve of XTE J1739$-$302 in the 2--10 keV band
 with 16~s bin.
 The vertical axis is the mean count rate of GIS~2 and GIS~3.
 The background is included, where the background level is
 $\sim$0.009 c~s$^{-1}$.
  The start time is MJD 51248.163188.
 \label{fig:XTEJ1739:lcur}}
\end{figure}

   We detected this source again with {\ASCA} on 1999 March 11 (Obs-ID=76).
 Figure~\ref{fig:XTEJ1739:lcur} shows the light curve in the
2--10 keV band.  Two flares, which started at around MJD 51248.319,
were detected.  

The light curve was unusual; it showed no flux, then suddenly flared up and reached the peak after  200--250~s, dropped to no flux level with  the same time
 scale as  the flare-rise (200--250~s).  The second flare had the almost identical profile except the peak flux of about a half of that of the first flare.  No significant difference in the light curves is found between  the
 hard (4--10 keV) and the soft (1--4 keV) energy bands.

\placefigure{fig:XTEJ1739:lcur}

We accumulated the spectrum only during the flare.  The spectrum is
 well fitted ($\chi^2$/dof = 91.4/77) with an absorbed power-law function of
 a photon index $\Gamma=0.80^{+0.10}_{-0.11}$ and a hydrogen column density
 {\NH}$=3.17^{+0.33}_{-0.31}\times 10^{22}$ {\NHUNIT}. 
 On the other hand, \citet{Smith1998_1739} reported that the {\XTE} spectrum
 on 1997 August 12 was well
 described with a thermal bremsstrahlung model of  $kT=12.4 \pm 0.3$ keV.
 Hence, the {\ASCA} result of the flat spectrum is very different from the {\XTE} result;
 we note that the power-law index obtained with {\ASCA} corresponds to a temperature over 100~keV
 when a bremsstrahlung model is applied.
The column density is also different from  the
 {\XTE}/PCA  result of (5--6)$\times 10^{22}$ {\NHUNIT} \citep{Smith1998_1739}.
 Thus, the spectral shape may have largely changed between the {\XTE} and the
 {\ASCA} observations.

We made the X-ray images in the pre-flare phase, and
found no  X-ray flux at the  source position with the 3$\sigma$ flux upper 
limit of 9$\times 10^{-13}$ {\FLUXUNIT} in the 2--10 keV band, assuming the same
 spectral shape during the flare phase.   Since the peak flux of the first flare is about 2$\times 10^{-9}$ {\FLUXUNIT}, this source exhibited drastic flux-increase by more than three orders within a few hundred seconds.
The  flat spectrum, highly variable flux, and absorption resemble
 those of transient pulsars.
We accordingly searched for pulsation in time scales of 1--100~s with
FFT analysis, and  epoch folding for time scales of 100--1000~s,
but found no significant pulsation, which is consistent with
 the {\XTE} result by \citet{Smith1998_1739}.

\subsection{SLX 1737$-$282 \label{sec:slx1737}}

The nature of SLX 1737$-$282 (Skinner {\etal} 1987), whether or not this is a persistent source,
 has been unknown.  
 We observed SLX 1737$-$282 (AX J1747.0$-$2818) five occasions from 1996 to 1999, and always detected positive flux. The position is determined to be
 ($l_{\rm II}$, $b_{\rm II}$)=(359$^{\circ}\!$.971, 1$^{\circ}\!$.231)
%($l_{\rm II}$, $b_{\rm II}$)=(359$^{\circ}\!$.9709, 1$^{\circ}\!$.2309)
 with an error radius of 40$''$, which is within the error radius of
 the {\Einstein} measurement (Skinner {\etal} 1987).  Therefore, we
 identified AX J1747.0$-$2818 as SLX 1737$-$282.

 The 2--10 keV flux of SLX 1737$-$282 was measured to be
 (3.5--5.0)$\times 10^{-11}$ {\FLUXUNIT} through the five {\ASCA} observations.
 Thus, SLX 1737$-$282 seems fairly stable in the long time span, but exhibited short time  variability of 10$^3$--10$^4$ seconds
 with amplitude of 100\% in the {\ASCA} observations.

The spectral shape had been also stable through 2.5-yr observations 
with a power-law function of a photon index $\Gamma=$ 2.1--2.4 and
a hydrogen column density  {\NH} $=$ (1.8--2.2)$\times 10^{22}$ {\NHUNIT}.
This column density implies that the distance of this source may be several kpc 
or farther.  The absorption-corrected
luminosity in the 2--10 keV band is (4--7)$\times 10^{35}$ {\LUMIUNIT}
for the assumed distance of 8.5~kpc.

The power-law index, moderate variability,  and the luminosity suggest 
that this source is a neutron star binary, although the other
possibilities, such as an AGN, are not totally excluded.

\subsection{GRS 1734$-$292 \label{sec:grs1734}}

GRS~1734$-$292 was discovered with {\Granat}/ART-P in 1990 by \citet{Sunyaev1990b}.
Since then, it has been observed in the wide energy band, from the soft X-ray band below
2~keV \citep{Barret_G1996} to the hard X-ray band
 \citep{Pavlinsky1994}, even up to 400~keV \citep{Churazov1992}. 
Mart\'{\i} {\etal} (1998)\markcite{Marti1998} found this source
to be a radio jet-like source, NVSS J173728$-$290802, and further identified
it to be a Seyfert~1 galaxy with $z=0.0214$, combining the radio, optical,
infrared, and X-ray results.

With {\ASCA}, the X-ray emission from this source has been always clearly detected
 whenever we observed this source; accordingly, GRS 1734$-$292 is probably
 a persistent X-ray source.  The X-ray spectrum was well represented with a power-law
function of a photon index of 1.3--1.7 and a hydrogen column density 
{\NH}$=$ (1.3--2.0)$\times  10^{22}$ {\NHUNIT}.

The power-law index of 1.3--1.7  is within the nominal range of that of Seyfert~1s.
The column density of (1.3--2.0)$\times  10^{22}$ {\NHUNIT} is significantly smaller than {\NH}$\sim 6\times 10^{22}$ {\NHUNIT} determined with {\Granat}/ART-P \citep{Pavlinsky1994}, but is consistent with the result of Mart\'{\i} {\etal} (1998)\markcite{Marti1998}, who estimated {\NH}$= (1.0\pm
0.2)\times 10^{22}$ {\NHUNIT} based on the measurement of optical
extinction and several absorption lines. 
 Therefore, our results support the Seyfert~1 identification by Mart\'{\i} {\etal} 
(1998)\markcite{Marti1998}.

\subsection{AX J1749.2$-$2725 \label{sec:axj1749}}

This source was discovered with {\ASCA} in 1995. \citet{Torii1998}
 found coherent pulsations of
 $P=220.38\pm 0.20$~s and $P=220.44\pm 0.80$~s periods in the Obs-ID=19, and Obs-ID=23 \& 24 observations, respectively.  The energy spectrum was described with
a flat power-law function (photon index of $\Gamma=$0.7--1.3)
with heavy absorption ({\NH}$=$(7--13)$\times 10^{22}$ {\NHUNIT})
through the {\ASCA} observations until 1997.

We  analyzed the data of Obs-ID=46 in 1998 March and 
confirmed the pulsations of  $P=219.9\pm 1.0$~s period.
Since the error  of the pulse period was large, we found
no significant $\dot{P}$ from the previous observations.
The spectral parameters except the flux were also  consistent in each observation.  The flux of $1\times 10^{-11}$ {\FLUXUNIT} was a few times smaller
than that in 1995 March (Obs-ID=19) and September (Obs-ID=23\&24), but 
a few times larger than that in 1996 September
(Obs-ID=33\&34) and 1997 September (Obs-ID=43).

\subsection{KS 1741$-$293 \& GRS 1741.9$-$2853 \label{sec:ks1741}}

KS 1741$-$293 and GRS 1741.9$-$2853 are X-ray bursters (in't Zand {\etal} 1991\markcite{Zand1991}; Cocchi {\etal} 1999a\markcite{Cocchi1999_1741.9}).
{\ASCA} found two sources at the position of 
($l_{\rm II}$, $b_{\rm II}$)=(359$^{\circ}\!$.5600, $-$0$^{\circ}\!$.0825)
and (359$^{\circ}\!$.9507, 0$^{\circ}\!$.1126) with the error radii of 40$''$. 
KS 1741$-$293 and GRS 1741.9$-$2853 lie within these error circles (e.g., in't Zand {\etal} 1991\markcite{Zand1991}, 1998\markcite{Zand1998};
Mandrou 1990\markcite{Mandrou1990};
Sunyaev 1990a\markcite{Sunyaev1990a}; Pavlinsky {\etal} 
1994\markcite{Pavlinsky1994}).

Although these two  are not persistent sources and although
we found  no burst from both the sources, 
the positional coincidence, the apparent brightness in their high states,
and the spectral parameters strongly favor that AX J1744.8$-$2921
and AX J1745.0$-$2855 (Table~\ref{tbl:res1:src-list}) are identical with KS 1741$-$293
and GRS 1741.9$-$2853, respectively.
We found that both the sources have showed apparent variability
by a factor of 50 or more.  The fluxes in the quiescent state were
less than a few 10$^{-12}$ {\FLUXUNIT} in the 0.7--10 keV band (see Table~\ref{tbl:res1:src-list}).

\acknowledgments

The authors express their thanks to the {\ASCA} team, and particularly,
the {\ASCA} Galactic plane/center survey team.
 We are grateful to Dr. K.~Torii and an anomymous referee for their valuable
 comments and suggestions.
 They appreciate the help of Dr. Y.~Ueda, Dr. Y.~Ishisaki, Mr. J.~Yokogawa,
 and Dr. L.~Angelini in the analysis.
 We acknowledge Mr. A.~Hands for correcting English.
 M. S. thanks Prof. K.~Nishikawa, Dr. H.~Awaki, and Dr. T.~Tsuru.
 M. S., H. M., and Y. M. are financially supported by
the Japan Society for the Promotion of Science for Young Scientists.
A part of this work was based on the data in the SIMBAD database.

%\addtocounter{figure}{-1}

%\begin{figure}[htbp]
% \centering
%% \psfig{file=imgsoft.eps,width=\textwidth,clip=}
%%% (b) 0.7--3.0 keV band.
%   \caption[f2b.eps]
%{ (continued from the previous page.)
%}
%\end{figure}

\clearpage

\begin{deluxetable}{rllrrrl}
% \tabletypesize{\footnotesize}
\tabletypesize{\scriptsize}
% \rotate
\tablecaption{Observation log \label{tbl:obs:log}}
\tablehead{
\colhead{Obs-ID} & \multicolumn{2}{c}{Date} & \multicolumn{2}{c}{Coordinates} & \colhead{Exp.\tablenotemark{a}} & \colhead{Target name} \\
\cline{2-3} \cline{4-5}
   & \colhead{Start} & \colhead{End} & \multicolumn{1}{c}{$l_{\rm II}$} & \multicolumn{1}{c}{$b_{\rm II}$} & & \\
   & \colhead{(UT)\tablenotemark{b}} & \colhead{(UT)\tablenotemark{b}} & \colhead{(deg.)} & \colhead{(deg.)} & \colhead{(ksec)} & 
% \cline{2-3} \cline{4-5}\\
}
\startdata
  1 & 1993/09/26-00:12 & 1993/09/27-05:08 &$-$1.02 & $-$0.08 &  30.5 &       1E1740.7$-$294 \\
  2 & 1993/09/30-17:35 & 1993/10/01-07:27 &$-$0.03 &    0.00 &  17.3 &              GC(0,0) \\
  3 & 1993/10/01-07:06 & 1993/10/01-21:06 &   0.31 &    0.21 &  17.7 &        GC(0,$-$0.37) \\
  4 & 1993/10/01-21:18 & 1993/10/02-10:14 &   0.49 & $-$0.09 &  17.2 &     GC(0.37,$-$0.37) \\
  5 & 1993/10/02-10:06 & 1993/10/03-00:58 &   0.15 & $-$0.30 &  19.9 &           GC(0.37,0) \\
  6 & 1993/10/03-00:30 & 1993/10/03-13:10 &$-$0.19 & $-$0.50 &  14.7 &        GC(0.37,0.37) \\
  7 & 1993/10/03-13:34 & 1993/10/04-02:58 &$-$0.29 & $-$0.15 &  20.3 &           GC(0,0.37) \\
  8 & 1993/10/04-02:54 & 1993/10/04-16:34 &$-$0.56 &    0.09 &  19.8 &     GC($-$0.37,0.37) \\
  9 & 1993/10/04-16:29 & 1993/10/05-03:21 &$-$0.22 &    0.30 &  17.9 &        GC($-$0.37,0) \\
 10 & 1994/09/08-19:40 & 1994/09/09-04:36 &$-$1.00 & $-$0.08 &  13.5 &   1E1740.7$-$2942 N1 \\
 11 & 1994/09/12-18:58 & 1994/09/13-01:06 &$-$0.99 & $-$0.09 &  11.8 &   1E1740.7$-$2942 N2 \\
 12 & 1994/09/15-22:46 & 1994/09/17-14:42 &$-$0.01 & $-$0.06 &  81.9 &                SGR A \\
 13 & 1994/09/22-03:08 & 1994/09/23-12:48 &   0.54 & $-$0.02 &  58.1 &             SGR B N1 \\
 14 & 1994/09/24-02:51 & 1994/09/24-14:11 &   0.54 & $-$0.02 &  19.7 &             SGR B N2 \\
 15 & 1995/03/15-08:22 & 1995/03/15-22:06 &   0.90 &    2.38 &  18.6 &       SLX 1735$-$269 \\
 16 & 1995/03/21-03:44 & 1995/03/22-00:04 &   0.03 & $-$0.84 &  25.1 &    GC(0min,$-$50min) \\
 17 & 1995/03/22-00:00 & 1995/03/22-22:48 &   0.03 & $-$1.25 &  27.0 &    GC(0min,$-$75min) \\
 18 & 1995/03/22-22:44 & 1995/03/23-20:36 &   0.03 & $-$1.67 &  25.3 &   GC(0min,$-$100min) \\
 19 & 1995/03/26-03:31 & 1995/03/27-12:55 &   1.48 &    0.28 &  35.6 & THE 200-PC RING N1 \\
 20 & 1995/03/27-12:50 & 1995/03/28-17:54 &$-$1.09 & $-$0.31 &  32.7 & THE 200-PC RING N2 \\
 21 & 1995/09/19-08:54 & 1995/09/21-21:38 &   0.31 &    0.03 &  63.0 &    THE 6.4KEV ISLAND \\
 22 & 1995/09/25-01:52 & 1995/09/25-08:36 &$-$1.83 & $-$0.08 &   8.6 &    GALACTIC RIDGE N1 \\
 23$^c$ & 1995/09/25-08:04 & 1995/09/25-14:08 &   1.83 &    0.16 &   5.0 &    GRO J1750$-$27 N1 \\
 24$^c$ & 1995/09/25-14:00 & 1995/09/25-17:20 &   1.51 &    0.15 &   2.5 &    GRO J1750$-$27 N2 \\
 25$^c$ & 1995/09/25-17:28 & 1995/09/26-00:44 &   2.16 &    0.25 &   8.9 &    GRO J1750$-$27 N3 \\
 26 & 1995/09/26-00:44 & 1995/09/26-09:00 &$-$2.33 & $-$0.08 &  12.7 &    GALACTIC RIDGE N2 \\
 27$^c$ & 1996/02/26-10:55 & 1996/02/27-02:47 &   0.17 &    0.27 &  14.0 &       GRO J1744$-$28 \\
 28 & 1996/03/08-17:19 & 1996/03/09-10:03 &$-$2.47 &    2.10 &  19.2 &                 HP 1 \\
 29 & 1996/09/10-03:56 & 1996/09/10-19:52 &$-$0.05 &    1.01 &  13.4 &              LB=0 75 \\
 30 & 1996/09/10-19:20 & 1996/09/11-07:52 &$-$0.03 &    1.58 &  18.8 &             LB=0 100 \\
 31$^d$ & 1996/09/11-07:47 & 1996/09/11-21:27 &$-$0.08 &    0.34 &  16.8 &              LB=0 50 \\
 32 & 1996/09/19-15:21 & 1996/09/19-22:33 &   0.98 &    0.00 &   8.2 &        GAL RIDGE2 N1 \\
 33 & 1996/09/19-21:29 & 1996/09/20-04:33 &   1.48 &    0.00 &  10.7 &        GAL RIDGE2 N2 \\
 34 & 1996/09/20-04:29 & 1996/09/20-09:21 &   1.97 &    0.00 &   7.7 &        GAL RIDGE2 N3 \\
 35 & 1996/09/20-08:45 & 1996/09/20-13:25 &   2.47 & $-$0.01 &   8.8 &        GAL RIDGE2 N4 \\
 36 & 1996/09/23-14:36 & 1996/09/24-02:32 &$-$0.90 & $-$0.51 &  14.8 &         G359.1$-$0.9 \\
 37 & 1996/09/28-05:19 & 1996/09/28-21:55 &$-$0.03 & $-$2.49 &  29.6 &           GC 0$-$150 \\
 38 & 1996/09/29-13:26 & 1996/09/30-17:46 &$-$0.03 & $-$2.07 &  36.8 &           GC 0$-$125 \\
 39$^c$ & 1997/03/16-15:57 & 1997/03/18-07:09 &$-$0.02 &    0.08 &  28.5 &              SGRA N1 \\
 40 & 1997/03/18-07:45 & 1997/03/18-17:41 &   0.78 &    1.16 &   9.3 &              SGRA N2 \\
 41 & 1997/03/20-21:36 & 1997/03/23-04:56 &$-$0.76 & $-$0.32 &  76.0 &         G359.1$-$0.2 \\
% 42$^e$ & 1997/04/02-06:21 & 1997/04/02-22:53 &$-$2.33 & $-$0.12 &  14.4 &        GRS 1737$-$31 \\
 42 & 1997/04/02-06:21 & 1997/04/02-22:53 &$-$2.33 & $-$0.12 &  14.4 &        GRS 1737$-$31 \\
 43 & 1997/09/15-11:57 & 1997/09/16-20:09 &   1.59 &    0.13 &   6.0 &     AGPS267.3$-$27.4 \\
 44$^e$ & 1997/09/19-05:28 & 1997/09/19-23:08 &$-$1.00 &    1.00 &  16.7 &           G359.1+0.9 \\
 45 & 1997/09/29-07:13 & 1997/09/30-12:49 &   0.56 &    1.20 &  32.3 &       GRS 1739$-$278 \\
 46 & 1998/03/13-21:05 & 1998/03/16-09:09 &   1.82 &    0.14 &   8.8 &         AGPS267.3 S1 \\
 47 & 1998/09/03-04:24 & 1998/09/03-10:00 &$-$1.01 & $-$1.01 &   6.1 &          GC REG 1 N1 \\
 48 & 1998/09/03-10:56 & 1998/09/03-20:44 &$-$0.51 & $-$1.02 &   9.7 &          GC REG 1 N2 \\
 49 & 1998/09/03-20:08 & 1998/09/04-03:16 &   0.49 & $-$1.01 &   9.1 &          GC REG 1 N3 \\
 50 & 1998/09/04-03:48 & 1998/09/04-09:44 &   0.99 & $-$1.01 &   8.3 &          GC REG 1 N4 \\
 51 & 1998/09/04-08:40 & 1998/09/04-16:08 &$-$0.51 & $-$0.52 &   8.4 &          GC REG 1 N5 \\
 52 & 1998/09/04-16:04 & 1998/09/05-00:40 &   0.24 & $-$0.52 &   8.4 &          GC REG 1 N6 \\
 53 & 1998/09/06-09:44 & 1998/09/06-18:44 &   0.74 & $-$0.52 &   5.7 &          GC REG 1 N7 \\
 54 & 1998/09/06-18:39 & 1998/09/06-21:07 &$-$0.75 &    0.48 &   3.3 &          GC REG 1 N8 \\
 55 & 1998/09/07-01:51 & 1998/09/07-07:35 &$-$0.26 &    0.49 &   7.4 &          GC REG 1 N9 \\
 56 & 1998/09/07-07:51 & 1998/09/07-15:27 &   0.24 &    0.48 &   9.3 &         GC REG 1 N10 \\
 57 & 1998/09/07-15:43 & 1998/09/07-23:07 &   0.74 &    0.48 &  10.0 &         GC REG 1 N11 \\
 58 & 1998/09/07-23:03 & 1998/09/08-05:59 &$-$0.51 &    0.98 &  10.0 &          GC REG 2 N1 \\
 59 & 1998/09/08-05:55 & 1998/09/08-11:43 &$-$0.76 &    1.48 &   7.9 &          GC REG 2 N2 \\
 60 & 1998/09/08-11:43 & 1998/09/08-21:51 &$-$0.26 &    1.48 &   9.5 &          GC REG 2 N3 \\
 61 & 1998/09/08-21:19 & 1998/09/09-03:43 &   0.24 &    1.48 &  10.0 &          GC REG 2 N4 \\
 62 & 1998/09/09-03:47 & 1998/09/09-18:59 &   0.50 &    0.98 &  14.5 &      UN SNR.5+1.0 N1 \\
 63 & 1998/09/09-18:27 & 1998/09/10-06:35 &   1.00 &    0.99 &  10.9 &      UN SNR.5+1.0 N2 \\
 64 & 1998/09/10-06:18 & 1998/09/10-14:06 &$-$1.02 &    1.99 &   8.1 &          GC REG 2 N5 \\
 65 & 1998/09/10-14:54 & 1998/09/10-22:58 &$-$0.53 &    1.99 &   9.7 &          GC REG 2 N6 \\
 66 & 1998/09/10-22:54 & 1998/09/11-05:42 &$-$0.02 &    1.99 &  10.0 &          GC REG 2 N7 \\
 67 & 1998/09/11-04:38 & 1998/09/11-11:58 &   0.98 &    1.99 &   8.2 &          GC REG 2 N8 \\
 68 & 1998/09/11-11:26 & 1998/09/11-19:58 &$-$1.42 &    0.00 &   7.2 &          GC REG 2 N9 \\
 69 & 1998/09/12-18:26 & 1998/09/12-23:34 &   0.71 &    1.49 &   7.4 &   UN SNR.75+1.5 1 N1 \\
 70 & 1998/09/12-23:18 & 1998/09/13-09:06 &   0.47 &    2.00 &  14.1 &   UN SNR.75+1.5 1 N2 \\
 71 & 1998/09/13-09:42 & 1998/09/15-02:50 &$-$0.83 & $-$0.81 &  60.3 &            THE MOUSE \\
% 72$^g$ & 1998/09/15-02:17 & 1998/09/15-09:57 &$-$0.55 &    0.00 &  10.1 &        THE MOUSE off \\
 72 & 1998/09/15-02:17 & 1998/09/15-09:57 &$-$0.55 &    0.00 &  10.1 &        THE MOUSE off \\
% 73$^h$ & 1998/09/26-03:26 & 1998/09/26-16:26 &   0.57 & $-$0.22 &  26.2 &      XTE J1748$-$288 \\
 73 & 1998/09/26-03:26 & 1998/09/26-16:26 &   0.57 & $-$0.22 &  26.2 &      XTE J1748$-$288 \\
 74 & 1999/03/10-16:09 & 1999/03/10-22:37 &$-$1.73 & $-$0.50 &  12.4 &           GC REG3 N1 \\
 75 & 1999/03/10-22:36 & 1999/03/11-03:56 &$-$1.23 & $-$0.49 &  10.3 &           GC REG3 N2 \\
 76 & 1999/03/11-03:45 & 1999/03/11-08:13 &$-$1.73 &    0.50 &  11.6 &           GC REG3 N3 \\
 77 & 1999/03/11-08:36 & 1999/03/11-14:04 &$-$1.23 &    0.51 &   8.8 &           GC REG3 N4 \\
 78 & 1999/03/11-14:32 & 1999/03/11-20:00 &$-$1.97 &    1.00 &  12.5 &           GC REG3 N5 \\
 79 & 1999/03/11-20:56 & 1999/03/12-01:32 &$-$1.47 &    1.00 &  10.3 &           GC REG3 N6 \\
 80 & 1999/03/12-01:56 & 1999/03/12-06:48 &$-$1.73 &    1.50 &   5.2 &           GC REG3 N7 \\
 81 & 1999/03/12-06:44 & 1999/03/12-12:48 &$-$1.23 &    1.51 &   6.1 &           GC REG3 N8 \\
 82 & 1999/03/12-12:16 & 1999/03/12-19:32 &$-$1.98 &    2.00 &  12.1 &           GC REG3 N9 \\
 83 & 1999/03/12-19:00 & 1999/03/13-00:40 &$-$1.48 &    2.00 &  10.8 &          GC REG3 N10 \\
 84 & 1999/03/14-15:48 & 1999/03/14-22:32 &   1.28 & $-$0.51 &  11.9 &         GC REG 4 N01 \\
 85 & 1999/03/14-22:35 & 1999/03/15-02:19 &   1.78 & $-$0.51 &   8.5 &         GC REG 4 N02 \\
 86 & 1999/03/15-02:03 & 1999/03/15-09:03 &   1.28 &    0.49 &   6.8 &         GC REG 4 N03 \\
 87 & 1999/03/15-09:27 & 1999/03/15-17:55 &   1.78 &    0.49 &   7.3 &         GC REG 4 N04 \\
 88 & 1999/03/15-16:47 & 1999/03/15-22:11 &   0.03 &    0.99 &   9.8 &         GC REG 4 N05 \\
 89 & 1999/03/15-21:35 & 1999/03/16-02:19 &   1.53 &    0.99 &  10.8 &         GC REG 4 N06 \\
 90 & 1999/03/16-02:43 & 1999/03/16-09:55 &   2.03 &    0.99 &   5.0 &         GC REG 4 N07 \\
 91 & 1999/03/16-08:47 & 1999/03/16-18:11 &   1.28 &    1.49 &   7.5 &         GC REG 4 N08 \\
 92 & 1999/03/16-18:35 & 1999/03/16-23:43 &   1.78 &    1.49 &  10.1 &         GC REG 4 N09 \\
 93 & 1999/09/25-03:02 & 1999/09/25-09:26 &$-$0.78 & $-$1.50 &   9.1 &          GC REG 5 N7 \\
 94 & 1999/09/25-09:22 & 1999/09/25-19:10 &$-$0.28 & $-$1.50 &   9.1 &          GC REG 5 N8 \\
 95 & 1999/09/25-19:06 & 1999/09/26-03:02 &$-$2.03 & $-$1.00 &   8.1 &          GC REG 5 N9 \\
 96 & 1999/09/26-02:26 & 1999/09/26-09:02 &$-$1.53 & $-$1.00 &   8.8 &         GC REG 5 N10 \\
 97 & 1999/09/29-13:04 & 1999/09/29-22:48 &   1.48 &    1.99 &   9.3 &         GC REG 4 N10 \\
 98 & 1999/09/29-22:56 & 1999/09/30-05:00 &   0.48 & $-$2.01 &   8.1 &          GC REG 6 N1 \\
 99 & 1999/09/30-05:16 & 1999/09/30-12:32 &   0.98 & $-$2.01 &   9.8 &          GC REG 6 N2 \\
100 & 1999/09/30-12:44 & 1999/09/30-22:24 &   1.48 & $-$2.01 &   9.5 &          GC REG 6 N3 \\
101 & 1999/09/30-22:20 & 1999/10/01-03:24 &   1.98 & $-$2.01 &  10.6 &          GC REG 6 N4 \\
102 & 1999/10/01-09:24 & 1999/10/01-20:32 &   0.23 & $-$1.51 &  10.2 &          GC REG 6 N5 \\
103 & 1999/10/01-21:32 & 1999/10/02-01:00 &   0.73 & $-$1.51 &   7.6 &          GC REG 6 N6 \\
104 & 1999/10/02-01:28 & 1999/10/02-06:56 &   1.23 & $-$1.51 &   6.6 &          GC REG 6 N7 \\
105 & 1999/10/02-07:28 & 1999/10/02-15:16 &   1.73 & $-$1.51 &   9.0 &          GC REG 6 N8 \\
106 & 1999/10/02-15:36 & 1999/10/02-23:00 &   1.48 & $-$1.01 &  10.4 &          GC REG 6 N9 \\
107 & 1999/10/02-23:56 & 1999/10/03-05:32 &   1.98 & $-$1.01 &   8.6 &         GC REG 6 N10 \\
\enddata

\tablenotetext{a}{The average exposure with GIS2 and GIS3 after the screening.}
\tablenotetext{b}{(Year/Month/Date-Hour:Minute).}
\tablenotetext{c}{GIS bit assignment of 10-8-8-0-0-5.} %% 23--25
% \tablenotetext{d}{GIS bit assignment of 10-8-8-0-0-5.} % 27
\tablenotetext{d}{GIS bit assignment of 8-8-8-5-0-2 for high and low bit-rate data, and 8-6-6-5-0-6 for medium bit-rate data.} % 31
% \tablenotetext{f}{GIS bit assignment of 10-8-8-0-0-5.} % 39
\tablenotetext{e}{GIS bit assignment of 8-8-8-5-0-2.} % 44
% \multicolumn{7}{l}{$^c$: TOO observations for GRO~J1750$-$27.  GIS bit assignment of 10-8-8-0-0-5.}\\ %% 23--25
% \multicolumn{7}{l}{$^d$: TOO observation for GRO~J1744$-$28.  GIS bit assignment of 10-8-8-0-0-5.}\\ % 27
% \multicolumn{7}{l}{$^e$: TOO observation for GRS~1737$-$31.}\\ % 42

\end{deluxetable}

\begin{deluxetable}{rlrrrrccllc}
% \tabletypesize{\footnotesize}
\tabletypesize{\scriptsize}
\rotate
\tablecaption{Point source list \label{tbl:res1:src-list}}
\tablehead{
% \colhead{Obs-ID} & \multicolumn{2}{c}{Date} & \multicolumn{2}{c}{Coordinates} & \colhead{Exp.\tablenotemark{a}} & \colhead{Target name} \\
 \colhead{Src-No\tablenotemark{a}} & \colhead{Name\tablenotemark{b}} &
 \colhead{$l_{\rm II}$\tablenotemark{c}} & \colhead{$b_{\rm II}$\tablenotemark{c}} & 
 \colhead{(Obs-ID)\tablenotemark{d}} & \colhead{$F_{\rm X}$\tablenotemark{e}} &
 \colhead{$\Gamma$\tablenotemark{f}} & \colhead{$N_{\rm H}$\tablenotemark{g}} &
 \colhead{ID} & \colhead{Category} &
 \colhead{Reference}
% \cline{2-3} \cline{4-5}\\
}
\startdata
  1 & AX J1734.5$-$2915 & $-$1.560 & 1.867 & 
         ( 83) & 3 & $\gtrsim 10$$^{}_{}$ & 11.0$^{+3.6}_{-<11.0}$ &  &  &  \\ %(631219_60_254_2:631219)
  2 & AX J1735.1$-$2930 & $-$1.696 & 1.642 & 
         ( 80) & 49 & 1.80$^{+0.73}_{-0.29}$ & $<0.9$ &  &  &  \\ %(631201_255_848_1:631201)
  3 & AX J1736.4$-$2910 & $-$1.258 & 1.571 & 
         ( 81) & 10 & 2.51$^{+1.9}_{-0.66}$ & $<1.5$ & HD 315992 & Star & 53 \\ %(631206_60_254_3:631206)
  4 & AX J1737.4$-$2907 & $-$1.111 & 1.419 & 
         ( 59) & 430 & 1.41$^{+0.52}_{-0.50}$ & 1.1$^{+1.1}_{-<1.1}$ & GRS 1734$-$292 & Seyfert-I & 1,3,33 \\ %(631206_255_848_1:590805)
 & & & & ( 81) & 310 & 1.51$^{+0.20}_{-0.19}$ & 1.7$^{+0.4}_{-0.3}$ &  &  &  \\ %(631206_255_848_1:631206)
  5 & AX J1738.2$-$2659 & 0.800 & 2.407 & 
         ( 15) & 1800 & 2.18$^{+0.03}_{-0.03}$ & 1.6$^{+0.1}_{-0.1}$ & SLX 1735$-$269 & Burster & 7,10,12,17 \\ %(231508_255_848_1:231508)
  6 & AX J1738.4$-$2902 & $-$0.917 & 1.280 & 
         ( 44) & 7 & 3.6$^{+>6}_{-1.3}$ & $<1.1$ & RX J1738.4$-$2901 &  & 53 \\ %(491905_60_254_2:491905)
 & & & & ( 59) & $<$13 & 3.6(fix)$^{}_{}$ & 0.0(fix)$^{}_{}$ &  &  &  \\ %(491905_60_254_2:590805)
  7 & AX J1739.1$-$3020 & $-$1.932 & 0.456 & 
         ( 76) & 1100 & 0.85$^{+0.10}_{-0.10}$ & 3.2$^{+0.3}_{-0.3}$ & XTE J1739$-$302 &  & 47 \\ %(631103_255_848_1:631103)
  8 & AX J1739.3$-$2923 & $-$1.117 & 0.926 & 
         ( 44) & 19 & 1.0$^{+1.6}_{-1.2}$ & 1.8$^{+6.0}_{-<1.8}$ &  &  &  \\ %(491905_255_848_3:491905)
  9 & AX J1739.4$-$2656 & 0.979 & 2.199 & 
         ( 15) & $<$6 & 6.3(fix)$^{}_{}$ & 13(fix)$^{}_{}$ &  &  &  \\ %(591105_255_848_2:231508)
 & & & & ( 67) & 7 & 6.3$^{+>4}_{-5.8}$ & 13$^{+19}_{-<13}$ &  &  &  \\ %(591105_255_848_2:591105)
 10 & AX J1739.5$-$2910 & $-$0.903 & 1.013 & 
         ( 44) & 4 & 3.9$^{+5.8}_{-1.3}$ & 0.2$^{+1.8}_{-<0.2}$ & ?\tablenotemark{h} & ?\tablenotemark{h} & 18,53 \\ %(491905_60_254_5:491905)
 11 & AX J1739.5$-$2730 & 0.519 & 1.882 & 
         ( 70) & 15 & 1.07$^{+0.98}_{-0.97}$ & 0.9$^{+2.1}_{-<0.9}$ &  &  &  \\ %(591223_255_848_2:591223)
 12 & AX J1740.1$-$3102 & $-$2.410 & $-$0.100 & 
         ( 26) & $<$4 & 1.53(fix)$^{}_{}$ & 6.2(fix)$^{}_{}$ & GRS 1737$-$31 & BHC & 9,49 \\ %(440206_255_848_1:292600)
 & & & & ( 42) & 1600 & 1.53$^{+0.06}_{-0.06}$ & 6.2$^{+0.3}_{-0.2}$ &  &  &  \\ %(440206_255_848_1:440206)
 13 & AX J1740.1$-$2847 & $-$0.507 & 1.084 & 
         ( 58) & 40 & 0.55$^{+0.66}_{-0.74}$ & 2.6$^{+3.2}_{-2.3}$ & AX J1740.1$-$2847 & Pulsar & 41 \\ %(590723_255_848_1:590723)
 14 & AX J1740.2$-$2903 & $-$0.725 & 0.931 & 
         ( 44) & 49 & 0.62$^{+0.64}_{-0.40}$ & 0.1$^{+1.2}_{-<0.1}$ & 2RXP J174015.5-290333 &  & 54 \\ %(491905_255_848_1:491905)
 & & & & ( 58) & 53 & $-$0.4$^{+5.0}_{-1.1}$ & $<45$ &  &  &  \\ %(491905_255_848_1:590723)
 15 & AX J1740.4$-$2856 & $-$0.609 & 0.958 & 
         ( 58) & 7 & 2.66$^{+1.3}_{-0.68}$ & $<0.5$ & 1RXS J174024.6$-$285706 &  & 50 \\ %(590723_60_254_1:590723)
 16 & AX J1740.5$-$3014 & $-$1.691 & 0.260 & 
         ( 22) & 25 & 3.4$^{+>7}_{-3.7}$ & 10$^{+30}_{-<10}$ & SAX J1740.5$-$3013 &  & 53 \\ %(631103_255_848_2:292501)
 & & & & ( 76) & $<$18 & 3.4(fix)$^{}_{}$ & 10(fix)$^{}_{}$ &  &  &  \\ %(631103_255_848_2:631103)
 17 & AX J1740.5$-$2937 & $-$1.173 & 0.582 & 
         ( 77) & 8 & 4.6$^{+>5}_{-1.4}$ & $<2.3$ & 1RXS J174034.3$-$293749 &  & 50 \\ %(631108_60_254_1:631108)
 18 & AX J1740.7$-$2818 & $-$0.029 & 1.231 & 
         ( 29) & 630 & 2.10$^{+0.11}_{-0.11}$ & 1.9$^{+0.2}_{-0.2}$ & SLX 1737$-$282 &  & 44 \\ %(391003_60_254_1:391003)
 & & & & ( 30) & 500 & 2.17$^{+0.16}_{-0.15}$ & 2.0$^{+0.2}_{-0.2}$ &  &  &  \\ %(391003_60_254_1:391019)
 & & & & ( 60) & 650 & 2.34$^{+0.21}_{-0.19}$ & 2.2$^{+0.3}_{-0.3}$ &  &  &  \\ %(391003_60_254_1:590811)
 & & & & ( 61) & 650 & 2.36$^{+0.31}_{-0.29}$ & 2.0$^{+0.5}_{-0.4}$ &  &  &  \\ %(391003_60_254_1:590821)
 & & & & ( 88) & 480 & 2.18$^{+0.19}_{-0.18}$ & 1.9$^{+0.3}_{-0.2}$ &  &  &  \\ %(391003_60_254_1:631517)
 19 & AX J1742.5$-$2845 & $-$0.201 & 0.668 & 
         (  9) & 21 & 4.9$^{+>5}_{-1.6}$ & $<1.1$ & 1RXS J174230.4$-$284504 &  & 50 \\ %(590701_60_254_1:0a0416)
 & & & & ( 29) & $<$14 & 4.9(fix)$^{}_{}$ & 0.5(fix)$^{}_{}$ &  &  &  \\ %(590701_60_254_1:391003)
 & & & & ( 31) & $<$33 & 4.9(fix)$^{}_{}$ & 0.5(fix)$^{}_{}$ &  &  &  \\ %(590701_60_254_1:391107)
 & & & & ( 55) & 13 & 4.9$^{+>5}_{-2.1}$ & 0.5$^{+2.0}_{-<0.5}$ &  &  &  \\ %(590701_60_254_1:590701)
 & & & & ( 88) & 31 & 3.7$^{+>6}_{-1.2}$ & 0.1$^{+2.6}_{-<0.1}$ &  &  &  \\ %(590701_60_254_1:631517)
 20 & AX J1742.6$-$3022 & $-$1.564 & $-$0.200 & 
         ( 22) & 31 & 0.2$^{+3.7}_{-<2}$ & 2.6$^{+48}_{-<2.6}$ &  &  &  \\ %(292501_255_848_1:292501)
 & & & & ( 68) & 23 & 1.9$^{+>8}_{-2.1}$ & 3.6$^{+45}_{-<3.6}$ &  &  &  \\ %(292501_255_848_1:591111)
 & & & & ( 74) & 54 & 0.9$^{+7.6}_{-<3}$ & 18$^{+79}_{-18}$ &  &  &  \\ %(292501_255_848_1:631016)
 21 & AX J1742.6$-$2901 & $-$0.414 & 0.498 & 
         ( 54) & $<$16 & 10(fix)$^{}_{}$ & 10.1(fix)$^{}_{}$ & 2RXP J174241.8-290215 &  & 54 \\ %(590701_60_254_2:590618)
 & & & & ( 55) & 7 & $\gtrsim 10$$^{}_{}$ & 10.1$^{+2.2}_{-5.4}$ &  &  &  \\ %(590701_60_254_2:590701)
 22 & AX J1743.9$-$2846 & $-$0.060 & 0.400 & 
         (  9) & 10 & 2.6$^{+1.7}_{-1.0}$ & 0.7$^{+1.2}_{-<0.7}$ & 2RXP J174351.3-284640 &  & 54 \\ %(431615_60_254_10:0a0416)
 & & & & ( 27) & 35 & $>$4.8 & 9.6$^{+2.5}_{-2.1}$ &  &  &  \\ %(431615_60_254_10:322610)
 & & & & ( 31) & 45 & 0.23$^{+0.34}_{-0.26}$ & $<0.6$ &  &  &  \\ %(431615_60_254_10:391107)
 & & & & ( 39) & 94 & 1.87$^{+1.3}_{-0.80}$ & 0.2$^{+0.5}_{-<0.2}$ &  &  &  \\ %(431615_60_254_10:431615)
 & & & & ( 55) & $<$8 & 1.87(fix)$^{}_{}$ & 0.2(fix)$^{}_{}$ &  &  &  \\ %(431615_60_254_10:590701)
 & & & & ( 56) & $<$22 & 1.87(fix)$^{}_{}$ & 0.2(fix)$^{}_{}$ &  &  &  \\ %(431615_60_254_10:590707)
 23 & AX J1743.9$-$2945 & $-$0.887 & $-$0.113 & 
         (  1) & 2100 & 1.36$^{+0.07}_{-0.07}$ & 13.1$^{+0.5}_{-0.5}$ & 1E 1740.7$-$2942 & BHC & 2,5,16,20,27,35,40 \\ %(432021_255_848_6:092600)
 & & & & ( 10) & 1600 & 1.51$^{+0.11}_{-0.11}$ & 13.6$^{+0.8}_{-0.8}$ &  &  &  \\ %(432021_255_848_6:190819)
 & & & & ( 11) & 1500 & 1.41$^{+0.12}_{-0.12}$ & 13.3$^{+0.9}_{-0.8}$ &  &  &  \\ %(432021_255_848_6:191218)
 & & & & ( 20) & 760 & 1.40$^{+0.30}_{-0.28}$ & 12.0$^{+2.1}_{-1.8}$ &  &  &  \\ %(432021_255_848_6:232712)
 & & & & ( 41) & 2500 & 1.13$^{+0.06}_{-0.06}$ & 12.2$^{+0.4}_{-0.4}$ &  &  &  \\ %(432021_255_848_6:432021)
 24 & AX J1744.5$-$2844 & 0.041 & 0.307 & 
         (  2) & $<$14 & 1.14(fix)$^{}_{}$ & 5.5(fix)$^{}_{}$ & GRO J1744$-$28 & Bursting pulsar & 14,29,36 \\ %(431615_255_848_1:093017)
 & & & & (  3) & $<$16 & 1.14(fix)$^{}_{}$ & 5.5(fix)$^{}_{}$ &  &  &  \\ %(431615_255_848_1:0a0107)
 & & & & (  9) & $<$12 & 1.14(fix)$^{}_{}$ & 5.5(fix)$^{}_{}$ &  &  &  \\ %(431615_255_848_1:0a0416)
 & & & & ( 12) & $<$15 & 1.14(fix)$^{}_{}$ & 5.5(fix)$^{}_{}$ &  &  &  \\ %(431615_255_848_1:191522)
 & & & & ( 27) & 200000 & 1.14$^{+0.01}_{-0.01}$ & 5.5$^{+0.1}_{-0.1}$ &  &  &  \\ %(431615_255_848_1:322610)
 & & & & ( 31) & 6000 & 0.99$^{+0.03}_{-0.03}$ & 5.4$^{+0.1}_{-0.1}$ &  &  &  \\ %(431615_255_848_1:391107)
 & & & & ( 39) & 50000 & 0.87$^{+0.01}_{-0.01}$ & 4.7$^{+0.1}_{-0.1}$ &  &  &  \\ %(431615_255_848_1:431615)
 & & & & ( 55) & $<$40 & 1.14(fix)$^{}_{}$ & 5.5(fix)$^{}_{}$ &  &  &  \\ %(431615_255_848_1:590701)
 & & & & ( 56) & $<$28 & 1.14(fix)$^{}_{}$ & 5.5(fix)$^{}_{}$ &  &  &  \\ %(431615_255_848_1:590707)
 25 & AX J1744.8$-$2921 & $-$0.440 & $-$0.083 & 
         (  7) & $<$23 & 2.12(fix)$^{}_{}$ & 20.3(fix)$^{}_{}$ & KS 1741$-$293 & Burster & 43,52 \\ %(591502_255_848_1:0a0313)
 & & & & (  8) & $<$24 & 2.12(fix)$^{}_{}$ & 20.3(fix)$^{}_{}$ &  &  &  \\ %(591502_255_848_1:0a0402)
 & & & & ( 72) & 910 & 2.12$^{+0.20}_{-0.19}$ & 20.3$^{+1.7}_{-1.6}$ &  &  &  \\ %(591502_255_848_1:591502)
 26 & AX J1745.0$-$2855 & $-$0.049 & 0.113 & 
         (  2) & $<$12 & 2.36(fix)$^{}_{}$ & 11.4(fix)$^{}_{}$ & GRS 1741.9$-$2853 & Burster & 6,32,37 \\ %(391107_255_848_10:093017)
 & & & & (  3) & $<$40 & 2.36(fix)$^{}_{}$ & 11.4(fix)$^{}_{}$ &  &  &  \\ %(391107_255_848_10:0a0107)
 & & & & (  7) & $<$20 & 2.36(fix)$^{}_{}$ & 11.4(fix)$^{}_{}$ &  &  &  \\ %(391107_255_848_10:0a0313)
 & & & & (  9) & $<$19 & 2.36(fix)$^{}_{}$ & 11.4(fix)$^{}_{}$ &  &  &  \\ %(391107_255_848_10:0a0416)
 & & & & ( 12) & 18 & 2.0$^{+2.1}_{-1.5}$ & 13.3$^{+14}_{-9.2}$ &  &  &  \\ %(391107_255_848_10:191522)
 & & & & ( 31) & 1000 & 2.36$^{+0.16}_{-0.16}$ & 11.4$^{+0.9}_{-0.8}$ &  &  &  \\ %(391107_255_848_10:391107)
 & & & & ( 39) & $<$8 & 2.36(fix)$^{}_{}$ & 11.4(fix)$^{}_{}$ &  &  &  \\ %(391107_255_848_10:431615)
% 27 & AX J1745.6$-$2901 & $-$0.079 & $-$0.043 & 
 27 & AX J1745.6$-$2901 & $-$0.079 & $-$0.044 & 
%         (  2) & 200 & 1.24$^{+0.79}_{-0.44}$ & 10.0$^{+7.1}_{-3.3}$ &  &  &  \\ %(0a0416_255_848_1:093017)
         (  2) & 200 & 1.24$^{+0.79}_{-0.44}$ & 10.0$^{+7.1}_{-3.3}$ & AX J1745.6$-$2901 & Burster & 24,30 \\ %(0a0416_255_848_1:093017)
 & & & & (  7) & 220 & 2.16$^{+0.95}_{-0.71}$ & 15.3$^{+9.1}_{-5.6}$ &  &  &  \\ %(0a0416_255_848_1:0a0313)
 & & & & ( 12) & 780 & 2.54$^{+0.11}_{-0.11}$ & 19.6$^{+0.9}_{-0.9}$ &  &  &  \\ %(0a0416_255_848_1:191522)
 & & & & ( 39) & 180 & 1.44$^{+0.62}_{-0.53}$ & 12.9$^{+6.0}_{-4.3}$ &  &  &  \\ %(0a0416_255_848_1:431615)
 28 & AX J1745.6$-$2900\tablenotemark{i} & $-$0.065 & $-$0.047 & 
         (  2) & 210 & 2.35$^{+0.37}_{-0.34}$ & 7.3$^{+1.3}_{-1.1}$ & Sgr A & MBH\tablenotemark{i} & 13,15,26 \\ %(191522_255_848_2:093017)
 & & & & (  7) & 170 & 2.66$^{+0.85}_{-0.70}$ & 8.4$^{+2.8}_{-2.2}$ &  &  &  \\ %(191522_255_848_2:0a0313)
% & & & & ( 12) & 380 & 1.80$^{+0.10}_{-0.10}$ & 7.5$^{+0.5}_{-0.5}$ &  &  &  \\ %(191522_255_848_2:191522)
 & & & & ( 12) & 320 & 2.35(fix) & 7.3(fix) &  &  &  \\ %(191522_255_848_2:191522)
% & & & & ( 21) & 200 & 2.62$^{+0.80}_{-0.67}$ & 9.0$^{+3.1}_{-2.4}$ &  &  &  \\ %(191522_255_848_2:291908)
% & & & & ( 27) & $<$190 & 1.80(fix)$^{}_{}$ & 7.5(fix)$^{}_{}$ &  &  &  \\ %(191522_255_848_2:322610)
 & & & & ( 39) & 240 & 2.17$^{+0.27}_{-0.25}$ & 7.5$^{+1.1}_{-0.9}$ &  &  &  \\ %(191522_255_848_2:431615)
 29 & AX J1746.1$-$2931 & $-$0.443 & $-$0.401 & 
         (  6) & 3100 & 1.40$^{+0.10}_{-0.10}$ & 5.9$^{+0.4}_{-0.4}$ & A 1742$-$294 & Burster & 4,28,50 \\ %(432021_255_848_1:0a0300)
 & & & & (  7) & 2700 & 1.64$^{+0.10}_{-0.09}$ & 6.5$^{+0.4}_{-0.4}$ &  &  &  \\ %(432021_255_848_1:0a0313)
 & & & & ( 41) & 6300 & 1.84$^{+0.03}_{-0.03}$ & 6.6$^{+0.1}_{-0.1}$ &  &  &  \\ %(432021_255_848_1:432021)
 & & & & ( 51) & 2500 & 1.82$^{+0.07}_{-0.06}$ & 5.8$^{+0.2}_{-0.2}$ &  &  &  \\ %(432021_255_848_1:590409)
 30 & AX J1746.3$-$2843 & 0.260 & $-$0.031 & 
         (  2) & 1100 & 2.11$^{+0.22}_{-0.21}$ & 19.5$^{+1.9}_{-1.7}$ & 1E 1743.1$-$2843 &  & 8,51 \\ %(291908_255_848_1:093017)
 & & & & (  3) & 1200 & 1.90$^{+0.21}_{-0.20}$ & 19.1$^{+1.9}_{-1.8}$ &  &  &  \\ %(291908_255_848_1:0a0107)
 & & & & (  4) & 1200 & 1.77$^{+0.19}_{-0.19}$ & 17.6$^{+1.6}_{-1.5}$ &  &  &  \\ %(291908_255_848_1:0a0121)
 & & & & (  5) & 1500 & 1.42$^{+0.18}_{-0.17}$ & 16.7$^{+1.5}_{-1.4}$ &  &  &  \\ %(291908_255_848_1:0a0210)
 & & & & ( 12) & 1300 & 1.83$^{+0.08}_{-0.08}$ & 18.2$^{+0.7}_{-0.7}$ &  &  &  \\ %(291908_255_848_1:191522)
 & & & & ( 13) & 1200 & 1.76$^{+0.13}_{-0.12}$ & 17.4$^{+1.1}_{-1.0}$ &  &  &  \\ %(291908_255_848_1:192203)
 & & & & ( 14) & 1500 & 1.71$^{+0.19}_{-0.19}$ & 18.9$^{+1.7}_{-1.6}$ &  &  &  \\ %(291908_255_848_1:192402)
 & & & & ( 21) & 710 & 2.12$^{+0.09}_{-0.09}$ & 18.6$^{+0.8}_{-0.8}$ &  &  &  \\ %(291908_255_848_1:291908)
 & & & & ( 39) & 910 & 1.65$^{+0.35}_{-0.33}$ & 20.6$^{+3.3}_{-3.0}$ &  &  &  \\ %(291908_255_848_1:431615)
 31 & AX J1747.0$-$2828 & 0.552 & $-$0.025 & 
         (  3) & $<$30 & 3.0(fix)$^{}_{}$ & 25(fix)$^{}_{}$ &  &  &  \\ %(192203_255_848_5:0a0107)
 & & & & (  4) & $<$12 & 3.0(fix)$^{}_{}$ & 25(fix)$^{}_{}$ &  &  &  \\ %(192203_255_848_5:0a0121)
 & & & & ( 13) & 8 & 3.0$^{+2.7}_{-1.7}$ & 25$^{+32}_{-15}$ &  &  &  \\ %(192203_255_848_5:192203)
 & & & & ( 14) & 8 & 1.9$^{+4.2}_{-2.3}$ & 9.4$^{+24}_{-<9.4}$ &  &  &  \\ %(192203_255_848_5:192402)
 & & & & ( 21) & $<$11 & 3.0(fix)$^{}_{}$ & 25(fix)$^{}_{}$ &  &  &  \\ %(192203_255_848_5:291908)
 & & & & ( 73) & 15 & 0.0$^{+1.9}_{-1.3}$ & 1.0$^{+6.5}_{-<1.0}$ &  &  &  \\ %(192203_255_848_5:592603)
 32 & AX J1747.0$-$2837 & 0.423 & $-$0.107 & 
         (  3) & $<$39 & 1.5(fix)$^{}_{}$ & 7.8(fix)$^{}_{}$ &  &  &  \\ %(192203_255_848_6:0a0107)
 & & & & (  4) & 12 & 1.3$^{+>9}_{-<3}$ & 8.8$^{+38}_{-<8.8}$ &  &  &  \\ %(192203_255_848_6:0a0121)
 & & & & (  5) & $<$39 & 1.5(fix)$^{}_{}$ & 7.8(fix)$^{}_{}$ &  &  &  \\ %(192203_255_848_6:0a0210)
 & & & & ( 13) & 10 & 1.5$^{+5.0}_{-<4}$ & 7.8$^{+20}_{-<7.8}$ &  &  &  \\ %(192203_255_848_6:192203)
 & & & & ( 14) & $<$9 & 1.5(fix)$^{}_{}$ & 7.8(fix)$^{}_{}$ &  &  &  \\ %(192203_255_848_6:192402)
 & & & & ( 21) & $<$11 & 1.5(fix)$^{}_{}$ & 7.8(fix)$^{}_{}$ &  &  &  \\ %(192203_255_848_6:291908)
 & & & & ( 73) & $<$17 & 1.5(fix)$^{}_{}$ & 7.8(fix)$^{}_{}$ &  &  &  \\ %(192203_255_848_6:592603)
 33 & AX J1747.1$-$2809 & 0.828 & 0.123 & 
%         ( 13) & 4 & $>$5.2 & 1.4$^{+0.4}_{-0.3}$ & SNR G0.9+0.1 &  & 22,34 \\ %(192203_60_254_2:192203)
         ( 13) & 4 & $>$5.2 & 1.4$^{+0.4}_{-0.3}$ &  &  & \\ %(192203_60_254_2:192203)
 & & & & ( 14) & $<$9 & 10(fix)$^{}_{}$ & 1.4(fix)$^{}_{}$ &  &  &  \\ %(192203_60_254_2:192402)
 & & & & ( 32) & $<$14 & 10(fix)$^{}_{}$ & 1.4(fix)$^{}_{}$ &  &  &  \\ %(192203_60_254_2:391915)
 34 & AX J1747.3$-$2809 & 0.867 & 0.070 & 
         ( 13) & 28 & 1.2$^{+3.1}_{-1.8}$ & 17$^{+28}_{-13}$ & SNR G0.9+0.1 & SNR & 22,34 \\ %(391915_255_848_11:192203)
 & & & & ( 14) & 15 & 5.0$^{+>5}_{-4.9}$ & 32$^{+64}_{-25}$ &  &  &  \\ %(391915_255_848_11:192402)
 & & & & ( 32) & 23 & 9.10$^{+>1}_{-7.3}$ & 100$^{+32}_{-80}$ &  &  &  \\ %(391915_255_848_11:391915)
 35 & AX J1747.4$-$3000 & $-$0.706 & $-$0.894 & 
         ( 47) & 240 & 2.33$^{+0.94}_{-0.76}$ & 3.2$^{+2.1}_{-1.6}$ & SLX 1744$-$299 & Burster & 37,38 \\ %(591309_255_848_11:590304)
 & & & & ( 48) & 1100 & 1.79$^{+0.17}_{-0.16}$ & 3.7$^{+0.4}_{-0.4}$ &  &  &  \\ %(591309_255_848_11:590310)
 & & & & ( 71) & 1700 & 1.86$^{+0.03}_{-0.03}$ & 4.3$^{+0.1}_{-0.1}$ &  &  &  \\ %(591309_255_848_11:591309)
 36 & AX J1747.4$-$3003 & $-$0.747 & $-$0.923 & 
         ( 47) & 2400 & 2.08$^{+0.14}_{-0.13}$ & 5.3$^{+0.4}_{-0.4}$ & SLX 1744$-$300 & Burster & 38,45 \\ %(591309_255_848_1:590304)
 & & & & ( 48) & 2800 & 1.98$^{+0.10}_{-0.10}$ & 5.5$^{+0.3}_{-0.3}$ &  &  &  \\ %(591309_255_848_1:590310)
 & & & & ( 71) & 3500 & 2.14$^{+0.02}_{-0.02}$ & 5.9$^{+0.1}_{-0.1}$ &  &  &  \\ %(591309_255_848_1:591309)
 37 & AX J1747.8$-$2633 & 2.288 & 0.824 & 
%         ( 90) & 56000 & 1.76$^{+0.04}_{-0.04}$ & 2.2$^{+0.1}_{-0.1}$ &  &  &  \\ %(631602_255_848_1:631602)
%         ( 90) & 56000 & 1.76$^{+0.04}_{-0.04}$ & 2.2$^{+0.1}_{-0.1}$ & GX 3+1 & Burster & 31 \\ %(631602_255_848_1:631602)
         ( 90) & 60000 & 1.76$^{+0.04}_{-0.04}$ & 2.2$^{+0.1}_{-0.1}$ & GX 3+1 & Burster & 31 \\ %(631602_255_848_1:631602)
 38 & AX J1748.0$-$2829 & 0.663 & $-$0.221 & 
         (  4) & $<$24 & 1.81(fix)$^{}_{}$ & 7.3(fix)$^{}_{}$ & XTE J1748$-$288 & BHC & 23,25,39,46 \\ %(590609_255_848_1:0a0121)
 & & & & ( 13) & 9 & 7.1$^{+>3}_{-5.0}$ & 71$^{+55}_{-45}$ &  &  &  \\ %(590609_255_848_1:192203)
 & & & & ( 14) & $<$21 & 1.81(fix)$^{}_{}$ & 7.3(fix)$^{}_{}$ &  &  &  \\ %(590609_255_848_1:192402)
 & & & & ( 32) & $<$64 & 1.81(fix)$^{}_{}$ & 7.3(fix)$^{}_{}$ &  &  &  \\ %(590609_255_848_1:391915)
 & & & & ( 53) & 650 & 1.81$^{+0.86}_{-0.79}$ & 7.3$^{+3.8}_{-3.3}$ &  &  &  \\ %(590609_255_848_1:590609)
 & & & & ( 73) & 14 & 2.3$^{+1.4}_{-1.1}$ & 5.9$^{+4.3}_{-3.1}$ &  &  &  \\ %(590609_255_848_1:592603)
 39 & AX J1748.3$-$2854 & 0.334 & $-$0.492 & 
         ( 52) & 7 & 4.6$^{+>5}_{-2.8}$ & 5.4$^{+8.8}_{-3.9}$ &  &  &  \\ %(590416_60_254_7:590416)
 40 & AX J1748.6$-$2957 & $-$0.538 & $-$1.087 & 
         ( 48) & 16 & 1.77$^{+1.0}_{-0.90}$ & 0.9$^{+1.4}_{-<0.9}$ & HD 316341 & Star & 53 \\ %(590310_255_848_4:590310)
 41 & AX J1748.7$-$2709 & 1.878 & 0.328 & 
         ( 23) & 28 & 3.2$^{+>7}_{-4.0}$ & 8.5$^{+42}_{-<8.5}$ & SNR G1.9+0.3 & SNR & 19 \\ %(531321_255_848_4:292508)
 & & & & ( 25) & 30 & 2.1$^{+5.1}_{-2.3}$ & 6.0$^{+22}_{-<6.0}$ &  &  &  \\ %(531321_255_848_4:292517)
 & & & & ( 34) & $<$55 & 3.2(fix)$^{}_{}$ & 10.2(fix)$^{}_{}$ &  &  &  \\ %(531321_255_848_4:392004)
 & & & & ( 43) & $<$150 & 3.2(fix)$^{}_{}$ & 10.2(fix)$^{}_{}$ &  &  &  \\ %(531321_255_848_4:491511)
 & & & & ( 46) & 29 & 3.2$^{+3.6}_{-2.0}$ & 10.2$^{+19}_{-7.4}$ &  &  &  \\ %(531321_255_848_4:531321)
 & & & & ( 87) & 47 & 1.3$^{+1.3}_{-1.6}$ & 3.7$^{+5.1}_{-<3.7}$ &  &  &  \\ %(531321_255_848_4:631509)
 42 & AX J1749.1$-$2733 & 1.585 & 0.051 & 
         ( 19) & $<$13 & 2.1(fix)$^{}_{}$ & 25(fix)$^{}_{}$ &  &  &  \\ %(391922_255_848_10:232603)
 & & & & ( 23) & $<$48 & 2.1(fix)$^{}_{}$ & 25(fix)$^{}_{}$ &  &  &  \\ %(391922_255_848_10:292508)
 & & & & ( 24) & $<$34 & 2.1(fix)$^{}_{}$ & 25(fix)$^{}_{}$ &  &  &  \\ %(391922_255_848_10:292514)
 & & & & ( 33) & 15 & 2.1$^{+4.7}_{-2.6}$ & 25$^{+57}_{-21}$ &  &  &  \\ %(391922_255_848_10:391922)
 & & & & ( 43) & 46 & $\lesssim$ $-$3$^{}_{}$ & 7.4$^{+119}_{-<7.4}$ &  &  &  \\ %(391922_255_848_10:491511)
 & & & & ( 46) & 59 & $<$9.9 & 3.3$^{+93}_{-<3.3}$ &  &  &  \\ %(391922_255_848_10:531321)
 43 & AX J1749.1$-$2639 & 2.358 & 0.505 & 
         ( 25) & 7300 & 0.46$^{+0.08}_{-0.07}$ & 3.2$^{+0.3}_{-0.2}$ & GRO J1750$-$27 & Pulsar & 11,42 \\ %(292517_255_848_1:292517)
 44 & AX J1749.2$-$2725 & 1.699 & 0.108 & 
         ( 19) & 290 & 0.76$^{+0.91}_{-0.50}$ & 15.4$^{+9.6}_{-4.5}$ & AX J1749.2$-$2725 & Pulsar & 48 \\ %(232603_255_848_1:232603)
 & & & & ( 23) & 180 & 0.10$^{+1.00}_{-1.0}$ & 4.5$^{+8.1}_{-<4.5}$ &  &  &  \\ %(232603_255_848_1:292508)
 & & & & ( 24) & 160 & 0.9$^{+5.0}_{-<3}$ & 19$^{+50}_{-<19}$ &  &  &  \\ %(232603_255_848_1:292514)
 & & & & ( 33) & $<$29 & 0.51(fix)$^{}_{}$ & 10.4(fix)$^{}_{}$ &  &  &  \\ %(232603_255_848_1:391922)
 & & & & ( 34) & $<$69 & 0.51(fix)$^{}_{}$ & 10.4(fix)$^{}_{}$ &  &  &  \\ %(232603_255_848_1:392004)
 & & & & ( 43) & $<$28 & 0.51(fix)$^{}_{}$ & 10.4(fix)$^{}_{}$ &  &  &  \\ %(232603_255_848_1:491511)
 & & & & ( 46) & 100 & 0.51$^{+0.73}_{-0.64}$ & 10.4$^{+7.0}_{-5.2}$ &  &  &  \\ %(232603_255_848_1:531321)
 45 & AX J1750.5$-$2900 & 0.489 & $-$0.952 & 
         ( 49) & 9 & 0.4$^{+>10}_{-1.2}$ & $<75$ & SAX J1750.8$-$2900 &  & 21 \\ %(590320_255_848_4:590320)
 46 & AX J1751.1$-$2748 & 1.587 & $-$0.451 & 
         ( 84) & $<$9 & 2.7(fix)$^{}_{}$ & 0.3(fix)$^{}_{}$ &  &  &  \\ %(631422_60_254_4:631415)
 & & & & ( 85) & 8 & 2.7$^{+>7}_{-1.4}$ & 0.3$^{+7.5}_{-<0.3}$ &  &  &  \\ %(631422_60_254_4:631422)
 47 & AX J1753.5$-$2745 & 1.912 & $-$0.891 & 
         (107) & 10 & 0.5$^{+>10}_{-<3}$ & 21$^{+1000}_{-<21}$ &  &  &  \\ %(6a0223_255_848_2:6a0223)
 48 & AX J1754.0$-$2929 & 0.467 & $-$1.854 & 
         ( 98) & 14 & 1.2$^{+4.9}_{-1.1}$ & 0.4$^{+9.2}_{-<0.4}$ &  &  &  \\ %(692922_255_848_1:692922)
 49 & AX J1754.2$-$2754 & 1.851 & $-$1.101 & 
         (106) & 93 & 3.7$^{+1.4}_{-1.2}$ & 4.5$^{+2.8}_{-2.3}$ &  &  &  \\ %(6a0223_255_848_1:6a0215)
 & & & & (107) & 110 & 2.54$^{+0.34}_{-0.31}$ & 2.1$^{+0.6}_{-0.5}$ &  &  &  \\ %(6a0223_255_848_1:6a0223)
 50 & AX J1754.5$-$2927 & 0.543 & $-$1.938 & 
         ( 98) & 3 & 7.6$^{+>2}_{-3.8}$ & 3.7$^{+4.4}_{-2.8}$ &  &  &  \\ %(692922_60_254_1:692922)
 51 & AX J1755.2$-$3017 & $-$0.090 & $-$2.482 & 
         ( 37) & 5 & 1.4$^{+4.6}_{-2.4}$ & 10$^{+36}_{-<10}$ &  &  &  \\ %(392805_255_848_1:392805)
 52 & AX J1755.7$-$2818 & 1.679 & $-$1.586 & 
         (105) & 12 & 0.3$^{+3.3}_{-<2}$ & $<11$ &  &  &  \\ %(6a0207_255_848_3:6a0207)
 53 & AX J1758.0$-$2818 & 1.937 & $-$2.021 & 
         (101) & 6 & 2.1$^{+>8}_{-2.6}$ & 17$^{+133}_{-17}$ &  &  &  \\ %(693022_255_848_2:693022)
\enddata
\tablenotetext{a}{Source No (Src-No).}
\tablenotetext{b}{Name designated.}
\tablenotetext{c}{Galactic coordinates.}
\tablenotetext{d}{Observation ID (Obs-ID) defined in Table~\ref{tbl:obs:log}.}
\tablenotetext{e}{Observed X-ray flux in the 0.7--10 keV band [$10^{-13}${\FLUXUNIT}] or 3$\sigma$ upper limit where the same spectral shape as that in their brightest state is assumed.}
\tablenotetext{f}{Photon index $\Gamma$, where $N(E) dE = E^{-\Gamma} dE$.}
\tablenotetext{g}{Hydrogen column density [10$^{22}$ H~cm$^{-2}$].}
\tablenotetext{h}{It may be HD 316072 or SNR G359.1+00.9.}
\tablenotetext{i}{This source is observed as spatially-extended plasma (see \S\ref{sec:sgra}).}
\tablecomments{Errors are in 90\% confidence.}
\tablerefs{
  [1]Barret \& Grindlay 1996 
  [2]Bouchet {\etal} 1991  
  [3]Churazov {\etal} 1992  
  [4]Churazov {\etal} 1995  
  [5]Churazov {\etal} 1996  
  [6]Cocchi {\etal} 1999a 
  [7]Cocchi {\etal} 1999b  
  [8]Cremonesi {\etal} 1999  
  [9]Cui {\etal} 1997  
  [10]David {\etal} 1997  
  [11]Dotani {\etal} 1995  
  [12]Elvis {\etal} 1992  
  [13]Falcke {\etal} 1999 
  [14]Finger {\etal} 1996  
  [15]Ghez {\etal} 1998  
  [16]Goldwurm {\etal} 1994  
  [17]Goldwurm {\etal} 1996  
  [18]Gray 1994 
  [19]Green \& Gull 1984 
  [20]Heindl {\etal} 1994  
  [21]Heise 1997  
  [22]Helfand \& Becker 1987 
  [23]Hjellming {\etal} 1998 
  [24]Kennea \& Skinner 1996
  [25]Kotani {\etal} 2000  
  [26]Koyama {\etal} 1996 
  [27]Kuznetsov {\etal} 1997  
  [28]Lewin {\etal} 1976  
  [29]Lewin {\etal} 1996  
  [30]Maeda {\etal} 1996  
  [31]Makishima {\etal} 1983
  [32]Mandrou 1990 
  [33]Mart\'{\i} {\etal} 1998 
  [34]Mereghetti {\etal} 1998 
  [35]Mirabel {\etal} 1992  
  [36]Nishiuchi {\etal} 1999  
  [37]Pavlinsky {\etal} 1994  
  [38]Predehl \& Kulkarni 1995   
  [39]Rupen {\etal} 1998  
  [40]Sakano {\etal} 1999a  
  [41]Sakano {\etal} 2000b  
  [42]Scott {\etal} 1997  
  [43]Sidoli {\etal} 1999 
  [44]Skinner {\etal} 1987  
  [45]Skinner {\etal} 1990  
  [46]Smith {\etal} 1998a  
  [47]Smith {\etal} 1998b  
  [48]Torii {\etal} 1998  
  [49]Ueda {\etal} 1997 
  [50]Voges {\etal} 1999  
  [51]Watson {\etal} 1981 
  [52]in't Zand {\etal} 1991  
  [53]SIMBAD 
  [54]ROSAT(2RXP)
}
%\tableline

\end{deluxetable}

\clearpage

\begin{deluxetable}{lccccccllc}
% \tabletypesize{\footnotesize}
\tabletypesize{\scriptsize}
\rotate
\tablecaption{Extended source list (including candidates) \label{tbl:res1:extended}}
\tablehead{
 \colhead{Name} & \colhead{Src-No\tablenotemark{a}} & \colhead{Size\tablenotemark{b}} &
 \colhead{$\Gamma$\tablenotemark{c}} & \colhead{$kT$\tablenotemark{d}} & \colhead{$N_{\rm H}$\tablenotemark{e}} & 
 \colhead{$F_{\rm X}$\tablenotemark{f}} & \colhead{X-ray Natures} & \colhead{Radio Natures} &
 \colhead{Refs.} \\
% \cline{2-3} \cline{4-5}\\
}
\startdata
 %\cutinhead{Radio supernova remnants (SNRs)\tablenotemark{g}.}
 \multicolumn{10}{c}{Radio supernova remnants (SNRs)\tablenotemark{g}.}\\
\tableline
 G359.0$-$0.9 &    & $8\times 20$ & \nodata & $0.4\pm 0.1$ & $1.5\pm 0.5$ & 24 & Thermal(Shell) & Shell-like & 1\\
 G359.1$-$0.5\tablenotemark{h} &    & $\sim$10 & \nodata & 0.2 \& 4 & $\sim$8 & 10 & Thermal(Center-filled) & Shell-like & 1,7\\
 G359.1$+$0.9\tablenotemark{i} & 10 & $\sim$4(?) & \nodata & $\sim$0.7 & $\sim$0 & 3 & Soft X-rays (extended?) & Composite & \\
% G0.0$+$0.0 & \multicolumn{7}{l}{not resolved from the Sgr A region.}\\
% G0.3$+$0.0 & \multicolumn{7}{l}{not resolved from 1E 1743.1$-$2847.}\\
 G0.9$+$0.1  & 34 & $<1.5$ & $\sim$1.5 & $\sim$2 & $\sim$10 & 20 & not resolved & Composite & \\ %(0.869137, 0.083753) radio
% G1.0$-$0.1 & \multicolumn{7}{l}{in the stray light from GX~3+1.}\\
% G1.4$-$0.1 & \multicolumn{7}{l}{in the stray light from GX~3+1.}\\
 G1.9$+$0.3  & 41 & $<1.5$ & $\sim$3 & $\sim$1 & $\sim$10 & 30  & not resolved & Shell-like & \\ %(1.871220, 0.326424) radio
 \cutinhead{X-ray diffuse structures newly discovered.}
 G359.4$+$0.0\tablenotemark{j} &  & $\sim$6 & $\sim$2 & \nodata & 12.6$\pm 3.4$ & 31 & XRN & The Sgr~C cloud &  4\\
 G359.95$-$0.05 &  & $2\times 3$ & \nodata & $\lesssim$1 \& $\sim$10 & $\sim$6 \& $\sim$12 & 200 & Thermal(Center-filled) & The Sgr~A region & \\
 G0.0$-$1.3     &  & $40\times 15$ & \nodata & 0.4--0.6 & 1.1--1.5 & 100 & Thermal(Center-filled) & \nodata & \\ 
 G0.1$-$0.1\tablenotemark{k} &  & $15\times 10$ & \nodata & $\lesssim$1 \& $\sim$10 & $\sim$4 \& $\sim$6 & 650 & Thermal \& 6.4-keV line & around the Radio Arc & \\ 
 G0.56$-$0.01\tablenotemark{l} & 31 & $<1.5$ & \nodata & 4--8 & 5--7 & 8 & Strong iron-K & \nodata & 5,6\\ 
 G0.7$-$0.1  &    & $\sim$7 & $\sim$2 & \nodata & 83$\pm 23$ & 25 & XRN & The Sgr~B2 cloud & 2,3\\ 
 G1.0$-$0.1  &  & $15\times 60$ & $\sim$3.5 & $\sim$1.0 & $\sim$0.8 & 50 & Thermal?(Center-filled) & \nodata & \\ 
\tableline
\enddata
\tablenotetext{a}{Src-No. defined in Table~\ref{tbl:res1:src-list},
 when the source is not spatially resolved with {\ASCA}.}
\tablenotetext{b}{Apparent X-ray size in unit of arcmin.}
\tablenotetext{c}{Photon index when the spectrum is fitted with a power-law function.}
\tablenotetext{d}{Temperature [keV] when the spectrum is fitted with a thin thermal model.}
\tablenotetext{e}{Hydrogen column density [$10^{22}$ H~cm$^{-2}$].}
\tablenotetext{f}{Observed flux with the 0.7--10 keV band [10$^{-13}$ erg~s$^{-1}$~cm$^{-2}$].}
\tablenotetext{g}{D. A. Green 1998: http://www.mrao.cam.ac.uk/surveys/snrs/ .}
\tablenotetext{h}{Two components for thermal emission are required. See \citet{Yokogawa2000} and \citet{Bamba2000} for detail.}
\tablenotetext{i}{
 If the central point-like component, AX J1739.5$-$2910, is due to a star HD~316072 (see Table~\ref{tbl:res1:src-list}),
 the flux of the diffuse component is crucially contaminated by the emission.}
\tablenotetext{j}{X-ray reflection nebula (XRN), Sgr C, which is characterized with a strong neutral iron line \citep{Murakami2001}.}
\tablenotetext{k}{Two components for thermal emission are required.  A part of emission is overlapped with the Radio Arc.}
\tablenotetext{l}{
 The X-ray emitting region, which was not resolved with {\ASCA}
 (see Src-No=31 in Table~\ref{tbl:res1:src-list} or 
 Sakano {\etal} 1999d\markcite{Sakano1999snr}),
 was recently found to be truly extended
 with the {\it Chandra} observation (Senda {\etal} 2001\markcite{Senda2001}).}

\tablerefs{
 1: Bamba {\etal} 2000\markcite{Bamba2000};
 2: Koyama {\etal} 1996\markcite{Koyama1996};
 3: Murakami {\etal} 2000b\markcite{Murakami2000b};
 4: Murakami {\etal} 2001\markcite{Murakami2001};
 5: Sakano {\etal} 1999d\markcite{Sakano1999snr};
 6: Senda {\etal} 2001\markcite{Senda2001};
 7: Yokogawa {\etal} 2000\markcite{Yokogawa2000} 
}

\end{deluxetable}

\begin{deluxetable}{lcl}
\tablecolumns{3}
\tablecaption{Multi-line fit to the Sgr~A diffuse plasma spectrum in 1993
 \label{tbl:sgra_spec}}
\tablehead{
\colhead{Param.} & \colhead{Best-fit Value} & \colhead{Comment}
%\colhead{Model parameters} & \colhead{Best-fit value} & \colhead{Unit} & \colhead{Comment}
% \cline{2-3} \cline{4-5}\\
}
\startdata

% \cutinhead{Continuum}
\multicolumn{3}{c}{Continuum}\\
\tableline
%%Normalization$^a$}                   & \multicolumn{2}{c}{8.4(6.9--11.0)$\times$10$^{-3}$}   \\
$F_{\rm x}$       		& 2.0 & Observed flux (10$^{-11}$ erg s$^{-1}$ cm$^{-2}$; 2--10 keV)\\
$L_{\rm x}$                	& 2.9 & Luminosity (10$^{35}$ erg s$^{-1}$; 2--10 keV)           \\
\tableline
%\multicolumn{4}{l}{--------Continuum------------}\\
$N_{\rm H}$	            	& 6.7 (5.9--7.7)  & Column density (10$^{22}$ H cm$^{-2}$) \\
$kT$				& 8.0 (5.4--14.7) & Temperature (keV)			\\
 \cutinhead{Emission (K$\alpha$) lines in unit of (10$^{-5}$ photon s$^{-1}$ cm$^{-2}$)}
Si  & 1.5 (0.3--2.7)		& (1.86~keV; He-like)                \\
S   & 4.3 (2.5--6.2)		& (2.45~keV; He-like)                \\
Ar  & 2.5 (0.1--3.8)		& (3.14~keV; He-like)                \\
Ca  & 2.5 (0.3--3.6)		& (3.90~keV; He-like)                       \\
Fe  & $<$2.7		& (6.4~keV; Neutral)                       \\
Fe  & 13 (8--17)		& (6.7~keV; He-like)                \\
Fe  & 3.2 ($<$6.7)		& (6.97~keV; H-like)                       \\
\tableline
$\chi^2$/d.o.f.            		& 114.5/124		& 	                \\
 \tableline
\enddata
\tablecomments{
 The spectrum is accumulated from the region indicated
 in Figure~\ref{fig:sgra-img} (see also text). 
The fit was made with the model of seven
Gaussian lines and an absorbed thermal bremsstrahlung
 for the energy band of 1.5--10 keV.
 Because of the limited statistics, each line energy was
fixed to the theoretical value, and the line width was fixed to zero. 
 Uncertainties are in 90\% confidence.
 Note that the plasma has an additional soft component below 1.5~keV.
}
\end{deluxetable}

\end{document}